\documentclass[12pt, preprint]{aastex}
 \usepackage{graphicx}
\usepackage{subfigure}
\usepackage{epsfig}
\begin{document}

\title{Mid-Infrared Spectral Diagnostics of Luminous Infrared Galaxies}
\author{A. O. Petric\altaffilmark{1}, L. Armus\altaffilmark{1}, J. Howell\altaffilmark{1}, B. Chan\altaffilmark{2}, J. M. Mazzarella\altaffilmark{2}, A. S. Evans\altaffilmark{3,4}, J. A. Surace\altaffilmark{1}, D. Sanders\altaffilmark{5}, P. Appleton\altaffilmark{6}, V. Charmandaris\altaffilmark{7,8}, T. Diaz Santos\altaffilmark{7}, D. Frayer\altaffilmark{4}, S. Lord\altaffilmark{2}, S. Haan\altaffilmark{1}, H. Inami\altaffilmark{1}, K. Iwasawa\altaffilmark{9}, D. Kim\altaffilmark{3}, B. Madore\altaffilmark{10},  J. Marshall\altaffilmark{1}, H. Spoon\altaffilmark{11}, S. Stierwalt\altaffilmark{1}, E. Sturm\altaffilmark{12}, V. U\altaffilmark{5}, T. Vavilkin\altaffilmark{13}, S. Veilleux\altaffilmark{14}}

\altaffiltext{1}{Spitzer Science Center, California Institute of Technology, 1200 E. California Blvd.  91125}
\altaffiltext{2}{Infrared Processing and Analysis Center, California Institute of Technology, MS 100-22, 770 South Wilson Ave., Pasadena, CA, 91125}
\altaffiltext{3}{University of Virginia}
\altaffiltext{4}{National Radio Astronomy Observatory}
\altaffiltext{5}{Institute for Astronomy, University of Hawaii, Honolulu, HI 96822}
\altaffiltext{6}{NASA Herschel Science Center, California Institute of Technology, 1200 E. California Blvd.  91125}
\altaffiltext{7}{Department of Physics, University of Crete, P.O. Box 2208, GR-71003, Heraklion, Greece}
\altaffiltext{8}{FORTH/Observatoire de Paris}
\altaffiltext{9}{ICREA Barcelona,Spain}
\altaffiltext{10}{Carnegie Observatories, 813 Santa Barbara Street, Pasadena, CA 91101}
\altaffiltext{11}{Department of Astronomy, Cornell University, Ithaca, NY 14953}
\altaffiltext{12}{MPE, Postfach 1312, 85741 Garching, Germany}
\altaffiltext{13}{Department of Physics and Astronomy, SUNY Stony Brook, Stony Brook, NY 11794}
\altaffiltext{14}{Astronomy Department, University of Maryland, College Park, MD 20742}

\begin{abstract}
We present a statistical analysis of the mid-infrared (MIR) spectra of 248 luminous infrared (IR) galaxies (LIRGs) which comprise the Great Observatories All-sky LIRG Survey (GOALS) observed with the Infrared Spectrograph (IRS) on-board the Spitzer Space Telescope. The GOALS sample enables a direct measurement of the relative contributions of star-formation and active galactic nuclei (AGN) to the total IR emission from a large sample of local LIRGs.

The AGN contribution to the MIR emission (f$_{\rm{AGN}}$) is estimated by employing several diagnostics based on the properties of the  [NeV], [OIV] and [NeII] fine structure gas emission lines, the 6.2 $\mu$m PAH and the shape of the MIR continuum. We find that 18\% of all LIRGs contain an AGN and that in 10\% of all sources the AGN contributes more than 50\% of the total IR luminosity. Summing up the total IR luminosity contributed by AGN in all our sources suggests that AGN supply $\sim12\%$ of the total energy emitted by LIRGs.  

The average spectrum of sources with an AGN looks similar to the average spectrum of sources without an AGN, but it has lower PAH emission and a flatter MIR continuum. AGN dominated LIRGs have higher IR luminosities, warmer MIR colors and are found in interacting systems more often than pure starbursts LIRGs. However we find no linear correlations between these properties and (f$_{\rm{AGN}}$). We used the IRAC colors of LIRGs to confirm that finding AGN on the basis of their MIR colors may miss $\sim 40$\% of  AGN dominated (U)LIRGs.

\end{abstract} 

\section{Introduction}

The Infrared (IR) Astronomical Satellite (IRAS) provided the first unbiased survey of the sky at mid-infrared (MIR) and far-infrared (FIR) wavelengths, giving us a comprehensive census of the IR emission properties of galaxies in the local Universe. IR number counts have been used to trace the importance of IR emission as a function of redshift, to explore star-formation and galaxy evolution \citep[]{flores99, gispert00, franceschini01, chary01, chary04, elbaz02, lagache03, marleau04, lefloc05, caputi06, magnelli09}.

To understand the origin of the observed IR emission, MIR diagnostic tools based on ISO \citep[for an exhaustive review see:][]{gec00} have been developed to study the roles played by star formation, AGN, and shocks (interaction-driven and wind-driven) in producing the observed IR emission. These diagnostics permit a direct mapping between IR number counts as a function of redshift/luminosity/galaxy type and the evolution of accretion and star-formation. 

The most basic diagnostics employed to estimate the AGN contribution to the MIR emission in individual galaxies are the ratios of high to low ionization fine-structure emission lines: [NeV] 14.3 $\mu$m/[NeII] 12.8 $\mu$m and [OIV] 25.9 $\mu$m/[NeII] 12.8 $\mu$m. The [NeV] 14.3 $\mu$m and [OIV] 25.9 $\mu$m lines trace high ionization gas. These methods have been used by e.g. \citet{lutz99, genzel98, verma03, armus06, sturm02, spoon07, armus07, farrah07}. The ionization potential of [NeV] is 96 eV. This is too high to be produced by OB stars, therefore its detection in the integrated spectrum of a galaxy usually indicates the presence of an AGN. This is not true for the [OIV] line because it takes only 55 eV to ionize O$^{++}$. Empirically it has been shown that emission line ratios of [NeV]/[Ne II] $\geq$ 0.75 and [OIV]/[NeII] $\geq 1.75$ indicate that more than 50\% of the nuclear MIR emission is produced by an AGN \citep[e.g.][and ref. within]{armus07}.

Additional diagnostics of the relative contribution of starbursts (SB) and AGN to the MIR luminosity are based on the dust properties, in particular PAH emission lines and the continuum emission arising from dust being heated by a SB or an AGN. Mid-infrared continuum emission in galaxies arises from a combination of ionized interstellar gas, evolved stellar population, non-thermal emission from radio sources, very small grains and PAHs. Empirically their respective contributions can be roughly distinguished from the shape of the SED \citep{laurent00}. Both theoretical models \citep[e.g.][]{pier92, nenkova02, granato07, levenson07} and observations \citep[e.g.][] {alonso01} show that the AGN radiation field can heat grains such that the dust continuum emission becomes prominent between 3 and 6 $\mu$m. Galaxies with an AGN tend to have low 6.2 $\mu$m PAH equivalent width (EQW) \citep[e.g][]{genzel98, lutz99, rigop99, tran01, sturm00, desai07, wu09} due to the presence of a significant hot dust continuum and also because the hard AGN photons may destroy the PAH molecules.

Spitzer IRS observations of normal nearby optically classified galaxies confirmed that the MIR diagnostics described above, diagnostics based on a combination of high to low ionization line ratios and PAH strengths are indeed very effective at determining the AGN versus starburst contribution to the IR in nearby galaxies. These studies such as the Spitzer Infrared Nearby Galaxy Survey (SINGS) were based on galaxies from a wide range of environments, Hubble Types and dust contents \citep{dale06}. 
Most recently \citet{goulding09} looked at an optically selected nearby ($D<15 $ Mpc) sample of galaxies with  IR luminosity between  $3 \times 10^{9} ~\rm{and}~ 2\times 10^{11} L_{\odot}$ and found that 27\% have AGN based on [NeV] detections. Excluding the nearby LIRG NGC1068, the [NeV]/[NeII] ratios  and the PAH EQWs of these sources suggest that only ~5\% of sources are AGN dominated. IRS studies of 24 nearby starbursts with luminosities between 10$^{9.75}$ and  10$^{11.6}$ at an average distance of 33 Mpc 
provided high and low resolution spectral templates for pure starbursts 
\citep{bers09,brandl06}. \citet{brandl06} also found the spectral continuum slope longward of $15 ~\mu$m can be used to discriminate between SB and AGN powered sources. These authors also found that in pure starbursts the PAH EQWs are independent of $L_{IR}$ but that the luminosity of the PAH feature scales with $L_{IR}$, in particular the 6.2 $\mu$m feature can be used to approximate the total IR of the starburst. 


%

%
Ultra Lumionus Infra-Red Galaxies (ULIRGs  L$_{\rm{IR}[8-1000\mu m]} \ge 10^{12} \rm{L}_{\odot}$) have also been extensively studied in the MIR regime. ULIRGs chosen primarily from the 1-Jy survey (Kim et al. 1998), the IRAS 2-Jy survey (Strauss et al 1992) and the First/IRAS radio-far-infrared sample of Stanford et al. 2000 with redshifts between 0.018 and 0.93 and IR luminosities in the range $10^{11.7} ~\rm{to}~ 10^{13.3}$ to L$_{\odot} $ were observed with the IRS by \citet[][]{armus04, armus06, armus07, desai07, farrah07}. These authors determine that around 42\% of ULIRGs have an AGN present as indicated by the detection of the [NeV] emission line,  that in ~20-40\% of ULIRGs the AGN dominates the MIR emission and that the 6.2 $\mu$m PAH EQW is anti-correlated with the  24 $\mu$m luminosity and with the  IRAS  25/60 $\mu$m color.
More recently \citet{veilleux09a} presented IRS observations of 74 ULIRGs and 34 Palomar-Green (PG) Quasars (QSOs). These authors derived adjustments to the MIR diagnostics to estimate the AGN contribution to the total bolometric luminosity not just to the the MIR luminosity. They estimated that the average AGN contribution to the bolometric luminosity of ULIRGs is on average 35-40\% and that it ranges from 15-35\% among cool optically classified HII like and LINER ULIRGs to 50-75\% among warm Seyfert 2 and Seyfert 1 ULIRGs . 

While nearby normal and starburst galaxies as well as ULIRGs have been studied in detail with IRS, MIR spectroscopic studies have only recently unveiled the source of IR emission in Luminous Infrared Galaxies (LIRGs). LIRGs ( L$_{\rm{IR}[8-1000\mu m]} \ge 10^{11} \rm{L}_{\odot}$, with L$_{\rm{IR}}$ as defined in \citet{sandMir96}) emit the bulk of their energy in the far-infrared. 
LIRGs account for $\geq$ 50\% of the total 24 $\mu$m galaxy population with $f_{24\mu m} \geq 80$ Jy and for $\sim$ 50\% of the comoving star formation density at z$\sim $1 \citep[e.g.][]{lefloc05, caputi06, magnelli09}. Unlike ULIRGs, LIRGs span the full range of galaxy interactions from non-merging spirals to late stage mergers, bridging the luminosity gap between SBs and QSOs. 

 \citet{alonso09b, pere10} combine optical and MIR spectroscopy to determine the excitation conditions in a sample of 15 local LIRGS with IR luminosities $L_{IR}~=~10^{11.5}-10^{11.59} ~L_{\odot}$. 
These authors find that most of the LIRGs have MIR spectroscopic properties similar to pure starbursts and that the integrated values show in general larger 6.2 $\mu$m PAH EQW than the nuclear spectra suggesting that star formation partially masks the nuclear activity. 
 
However, until recently there has been no comprehensive study of the MIR spectra of a large sample of LIRGs. This paper employs MIR diagnostics aimed at isolating the AGN contribution to the IR emission in the study of 248 LIRGs. The Great Observatories All-sky LIRG Survey (GOALS)  targets a complete sample of 202 systems in the local Universe selected from the IRAS Revised Bright Galaxy Sample \citep[RBGS:][] {sanders03}. GOALS brings together HST (UV, optical and NIR), Spitzer (imaging and spectroscopy), GALEX, Chandra, and NIR ground-based spectroscopic data to understand low-z ($\leq $ 0.088) LIRGs. The broad wavelength coverage of this study can be used to trace the young and old stars, the dust, the hot ionized gas, and the warm molecular gas in LIRGs, making this data set unmatched in its power to shed light on the genesis of starbursts and the growth of black holes in interacting galaxies locally. An outline of the GOALS project and a multi-wavelength analysis of the LIRG VV340 are given in \citet{armus09}. \citet{evans08} and  \citet{inami10} present a similar multi-wavelength analysis of LIRGs NGC 2623 and CGCG448-020 respectively. Howell et al. (2009) compares the IR and UV properties of LIRGs and Diaz-Santos (2010) discusses the extent of the MIR emission across a large sample of LIRGs. 

A key component of GOALS is the mid-infrared spectroscopic study of 248 LIRG nuclei (in 202 systems) with the Infrared Spectrograph (IRS) on Spitzer.
In the present paper we discuss the first statistical results from GOALS on the MIR nuclear spectra focusing on the ionization source in LIRG nuclei at low redshift. In section 2 we describe the IRS observations and analysis. In section 3 we present the results based on key starburst and AGN emission line features that are used to diagnose the origin of the IR emission as well as statistics on the presence and strength of these features in the nuclei of LIRGs. In section 3 we also present an analysis of the properties of average low-resolution spectra obtained from samples of LIRGs without and with an AGN.  In sections 4.1 through 4.3 we present an investigation of how these diagnostics relate to the global properties of LIRGs. In section 4.4 we discuss the connections between AGN and merger activity in LIRGs.

In section 4.5 we compare spectroscopic diagnostics of AGN to diagnostics using only IR photometry. 
In the conclusion we summarize the findings of this work. This is the first in a series of papers which compare the MIR spectral and multi-wavelength properties of LIRGs in the local universe. 

\section{IRS Observations and Data Analysis}
Spectra of 248 individual nuclei in 202 LIRG systems, observed in all four IRS modules (Short-Low, Long-Low, Short-High, and Long-High), were used for this investigation. The widths of the SL, SH, LL, LH slits (3.6\arcsec , 4.7\arcsec, 10.7\arcsec, 11.1\arcsec) correspond to 1.5, 2.0, 4.5, 4.6 kpc respectively at a distance of 88 Mpc (the median galaxy distance of our sample). As part of GOALS, IRS spectra for 158 LIRG systems were obtained \citep[PID 30323][]{armus09}. Of these, 115 were observed in all four IRS modules, while 43 have been observed in three or fewer IRS modules in order to complete the existing archival data and to ensure complete coverage for all GOALS targets. Spectra for the remaining 44 LIRGs systems were obtained from other Spitzer IRS programs in the archive. In all data from PID 30323, IRS Staring Mode was employed, using ``cluster target'' observations for those sources with well separated ($\Delta{r} \geq 10\arcsec$), nearby interacting companions. Among the 202 LIRGs studied, secondary nuclei were targeted only when the flux ratio of primary to secondary nucleus (as measured in the MIPS 24 $\mu$m data) is less than or equal to five, in order to capture the spectra of the nuclei actively participating in the far-infrared emission of the system. The IRS data on two LIRGs (IIIZw35 and NGC 1614) in the GOALS sample do not include spectra of the nuclear regions, and as such they are not included in the statistics presented bellow. 

All data were reduced using the S15, S16 and S17 IRS pipelines at the Spitzer Science Center\footnote{$\rm{http://ssc.spitzer.caltech.edu/irs/features.html}$}. The pipelines were changed to modify certain header keywords, produce new flat-fields, and lower SL and LL fringes by 1-20\%, provide better treatment of ramp slopes, and make several changes to the post-bcd products which we did not use for this analysis. The changes made between the pipelines should not systematically alter the measurements presented here. The IRS pipeline includes ramp fitting, dark sky subtraction, droop correction, linearity correction and wavelength and flux calibration.
 
The backgrounds in the high resolution data were subtracted for all objects with dedicated sky observations ($\sim$ 60\% of the sources). For the low-resolution data, and for high resolution data without dedicated background observations, off-source nods were used for sky subtraction. Large objects in PID 30323 had dedicated background pointings for both low and high resolution data. Bad pixel mask files were combined such that the final masks flagged all the individual bad pixels, if they were marked as bad in one individual exposure. Basic calibrated data sets (BCDs) for each nod were combined by determining the median if more than 5 BCDs were available, otherwise the average was used with 3-sigma clipping. 
 
Each nod was extracted with SPICE\footnote{http://ssc.spitzer.caltech.edu/postbcd/doc/spice.pdf} using the standard extraction aperture and point-source calibration. Nod 1 and nod 2 were compared with each other, and pixels were flagged if the difference between nod 1 and nod 2 exceeded 30\%, and when adjacent pixels within the same nod differed by more than 30\% (due to a cosmic ray or hot pixel). Spatial profiles were computed from the SL data at 8.6, 10, and 12.8 $\mu$m and discussed in detail in the GOALS IRS delivery documents\footnote {$\rm{http://irsa.ipac.caltech.edu/data/SPITZER/GOALS}$}. From these data it was assessed that 20\% of the sources have extended emission at 12.8 $\mu$m while the remaining 80\% are point sources. 

Twenty eight systems were observed in spectral mapping mode. The 2D BCDs were assembled, then cleaned to remove obvious spikes, and finally the nuclear spectra were extracted with CUBISM \citep{smith07} using extraction regions of sizes equal to those of extraction regions for point sources in the spectra taken in staring mode. In total 36 nuclear spectra were extracted from data taken in spectral-mapping mode. 

Gaussian fits were performed on the high resolution data to measure the fluxes for each of the gas emission lines, and to deblend the [ClII] 14.368 $\mu$m -- [NeV] 14.322 $\mu$m lines and the [OIV] 25.890 $\mu$m -- [FeII] 25.988 $\mu$m lines. To properly estimate the 6.2 $\mu$m PAH EQW, the associated continuum was measured by using a spline fit to the continuum but excluding points affected by water ice absorption between 5.55 and 6.0 $\mu$m and hydrocarbon absorption. The amount of water ice and hydrocarbon absorption was estimated as described in \citet{spoon07}. All the spectra were inspected by eye to determine which sources had water ice absorption. For those sources with water ice absorption (4\% of all spectra) the continuum was estimated by using a linear interpolation of all data points in the following wavelength regions: [5.1 $\mu$m -5.6 $\mu$m], [6.7 $\mu$m-6.9 $\mu$m,], [7.1 $\mu$m -7.2 $\mu$m].  

For each spectrum the 5.5 $\mu$m flux was measured by integrating the continuum between 5.3 and 5.8 $\mu$m, and the 15 $\mu$m flux was estimated by integrating continuum between 14 and 15 $\mu$m. The nuclear 24 $\mu$m flux was calculated from the IRS LL spectrum using the spectral response of the MIPS 24 $\mu$m bandpass. The total 24 and 70 $\mu$m fluxes were measured from MIPS observations (Mazzarella et al.  in prep.) using typical apertures of 1 arcminute.

\section{Results}
The high resolution (SH and LH) LIRG spectra are dominated by atomic fine-structure lines of Ne, O, Si, S as well as warm H$_{2}$. Several methods to determine the relative contribution of star formation versus AGN to the IR luminosity of LIRGs using MIR spectra are used.
The results section is organized as follows: the ionized  gas diagnostics of AGN activity based on high and low ionization fine structure lines are discussed in section 3.1; the results arising from the dust diagnostics are presented in section 3.2; low resolution spectra are divided into four groups based on the presence and importance of an AGN to the nuclear MIR emission as determined from the ionized gas diagnostics discussed in section 3.1; the low resolution spectra of each group of sources are then averaged and these average spectra are presented in section 3.3. The properties of these average spectra and their implications for statistical studies of AGN activity are also briefly discussed in section 3.3. All wavelengths mentioned here are rest-frame values.

\subsection{Ionized gas diagnostics}


The [NeV] 14.3 $\mu$m line is detected in 43 nuclei representing 18\% of the GOALs sample.  Because detecting  [NeV] emission over kpc scales provides direct evidence for the presence of an AGN, the [NeV] properties can be directly compared with other direct indicators of AGN activity (e.g. those based on X-ray emission; Iwasawa et al. 2010). Table 1 lists all the sources with [NeV] detections at 14.3 $\mu$m as well as their [NeV]/[NeII] ratios and the ratio $L_{[NeV]}/L_{IR}$. The detected [NeV] fluxes range between $1.5 \times 10^{-18} \rm{W ~m^{-2}}$ and   $9 \times 10^{-15} \rm{W ~m^{-2}}$  with a median flux of $2.27 \times 10^{-17} \rm{W ~m^{-2}}$. The [NeV]/[NeII] ratios range between  $\sim 1\times 10^{-3}$ and 2.09, with a median of  0.07. The $L_{[NeV]}/L_{IR}$ range between $\sim 1\times 10^{-5} $ and $\sim 1\times 10^{-4}$. Note that the MIR spectra and [NeV] fluxes of the ULIRGs in GOALS have already been published in \citep{armus04,armus06,armus07,farrah07,veilleux09a}. 

\begin{figure}[htb]
   \begin{center}
   \includegraphics[width=\linewidth]{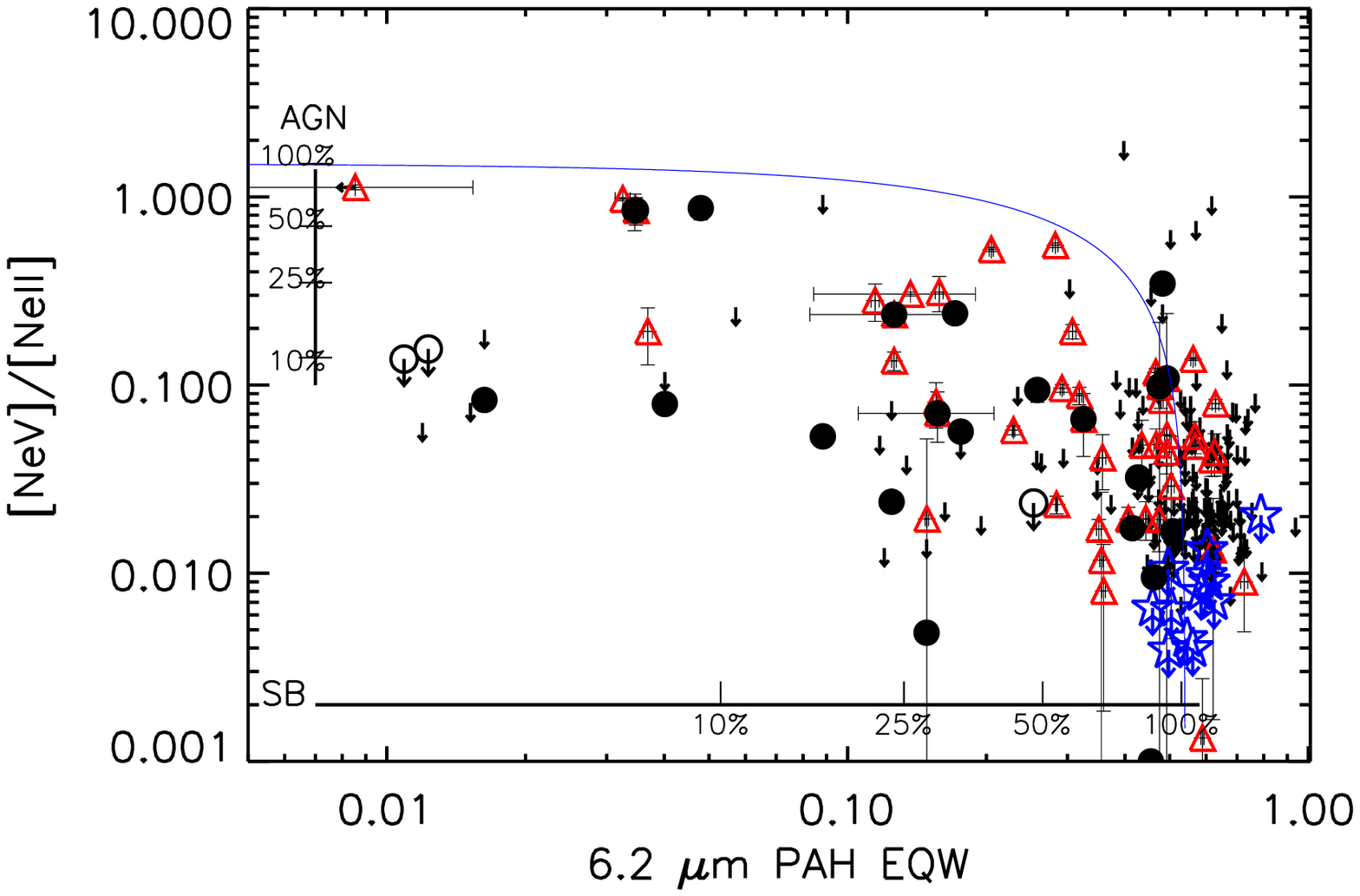} 
   \caption{Mid-infrared [NeV]/[NeII] versus 6.2 $\mu$m PAH EQW excitation diagram. The \textit{red triangles} are [NeV] detections while the \textit{black arrows} are upper limits. In all cases 1$\sigma$ error bars are shown.The black circles indicate ULIRGs where the filled symbols show detections and empty symbols upper limits. The blue empty stars mark upper limits for star-burst galaxies from \citet{bers09}. The \textit{solid black lines} indicate the fractional AGN and starburst contribution to the MIR luminosity from the [NeV]/[NeII] (vertical) and 6.2 $\mu$m PAH EQW (horizontal) assuming a simple linear mixing model. In each case, the 100$\%$, 50$\%$, 25$\%$, and 10$\%$ levels are marked. The 100$\%$ level is set by the average detected values for the [NeV]/NeII] and 6.2 $\mu$m EQW among AGN and starbursts respectively, as discussed in \citet{armus07}. The \textit{blue line} traces where the summed SB and AGN contribution equals 100$\%$. For most LIRGs the [NeV]/[NeII] ratio suggests that the AGN contribution to the nuclear MIR luminosity is below 10$\%$. }
   \label{Nov01_ne5_diag0}
   \end{center}
   \end{figure}

\begin{figure}[htb]
   \begin{center}
   \includegraphics[width=\linewidth]{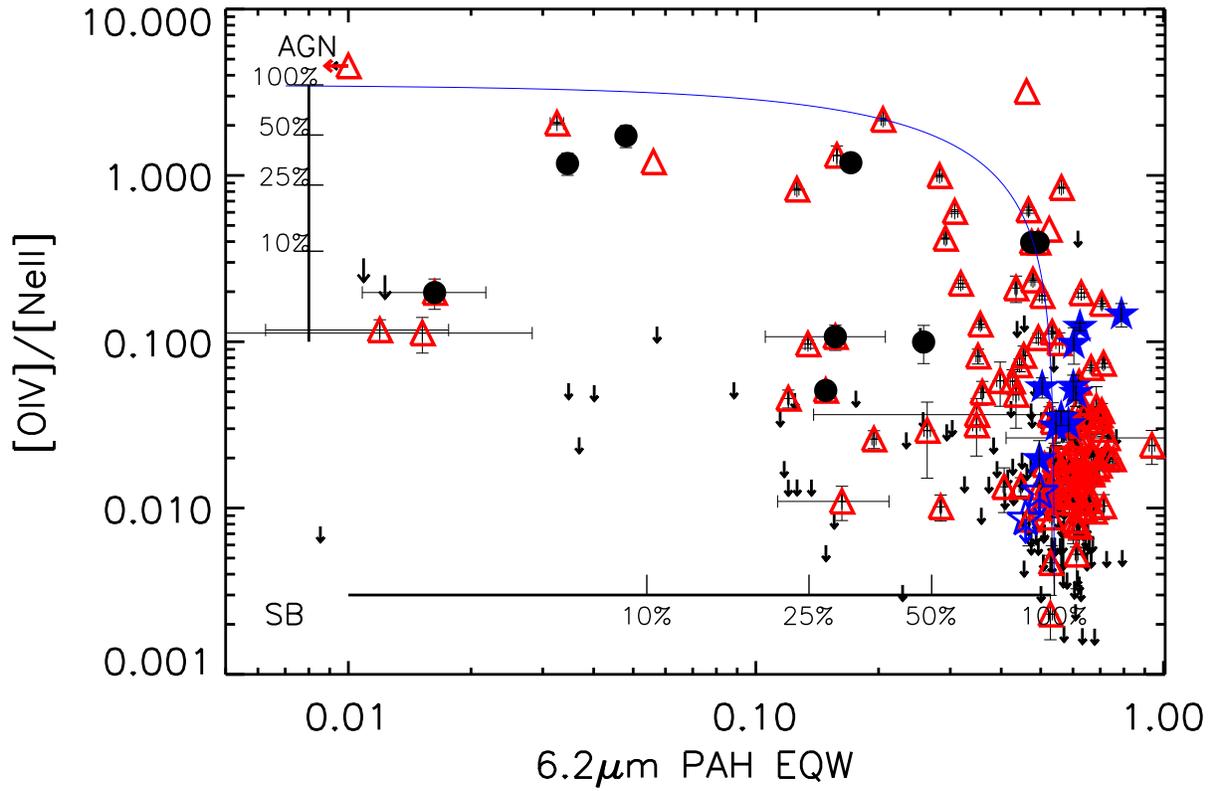}
   \caption{Mid-infrared [OIV]/[NeII] versus 6.2 $\mu$m PAH EQW excitation diagram. Symbols have the same definition as in Figure 1 except that red triangles indicate [OIV] detections}
   \label{O4ne2_basic_diag}
   \end{center}
   \end{figure}

Figures \ref{Nov01_ne5_diag0} and \ref{O4ne2_basic_diag} show the ratios of [NeV]/[NeII] and [OIV]/[Ne II] versus the EQW of the 6.2 $\mu$m PAH feature for the LIRGs in our sample. As mentioned in the introduction, empirically it has been shown that emission line ratios of [NeV]/[Ne II] $\geq$ 0.75 and [OIV]/[NeII] $\geq 1.75$ indicate that more than 50\% of the nuclear MIR emission is produced by an AGN \citep[e.g.][and ref. within]{armus07}. Three LIRG nuclei have [NeV]/[NeII] $\geq$ 0.75, implying that only $1\%$ of LIRGs have more than 50$\%$ of their MIR emission powered by an AGN (see Figure \ref{Nov01_ne5_diag0}).

 For comparison with the LIRGs, nine starburst galaxies are included in the figures using data from \citet{bers09}: NGC 660, NGC 1222, IC 342, NGC 1614, NGC, 2146, NGC 3256, NGC 3310, NGC 4088, NGC 4676, NGC 4818, NGC 7252, NGC 7714. Three of these galaxies are LIRGs while the rest have lower IR luminosities. These particular objects were used by \citet{armus07} to determine the zero points in the ionized gas diagnostics. Data for seven well studied ULIRGs from the RBGS sample analyzed in \citet{armus04,armus07} are also presented for comparison: Mrk 231, Arp 220, IRAS 05189-2524, Mrk 273, IRAS 08572+3915, UGC 5101, Mrk 1014. 

The [OIV] 25.890 $\mu$m line is detected in 120 nuclei (53\% of the sources). Four LIRGs have ratios of [OIV]/[NeII] $\geq 1.75$ implying that only $1\%$ of LIRGs have more than 50$\%$ of their MIR emission powered by an AGN. The ratios of [OIV]/[NeII] range between 0.002 and 5.5 and have a median of 0.03 and a mean of 0.24 with a dispersion of 0.74 (see Figure \ref{O4ne2_basic_diag}).

For several sources, as observed in similar investigations \citep[e.g.][]{farrah07}, the estimated contributions to the MIR luminosity from star-formation and an AGN do not add up to 100\%. In figures \ref{Nov01_ne5_diag0} and \ref{O4ne2_basic_diag} these sources are located either below or above the mixing line between the diagnostics, that is as the PAH EQW drops the [NeV]/[NeII] should go up, and galaxies should follow this line. However many sources fall well bellow this mixing line. We propose that this is partially due to uncertainties in the AGN zero points marking at what [NeV]/[NeII] and [OIV]/[Ne II] ratios an AGN contributes 100\% of the MIR luminosity. In addition differential extinction of the [NeV] and [OIV] lines originating from the obscured region around the nucleus versus the [NeII] line originating from the starburst (likely more extended) may lead to an underestimation of the contribution of the AGN to the MIR emission. 

\subsection{Dust diagnostics}
For the sample of LIRGs in GOALS the 6.2 $\mu$m PAH EQWs range between 0.01 and 0.94 $\mu$m with a mean of 0.47 and a median of 0.53. About 16\% of LIRG nuclei have 6.2 $\mu$m PAH EQW below 0.27 $\mu$m, less than one half of the value seen in local starburst systems \citep{brandl06}. This suggests that 16\% of LIRGs have an AGN which dominates the MIR emission.

\begin{figure}[htb]
   \begin{center}
   \includegraphics[width=\linewidth]{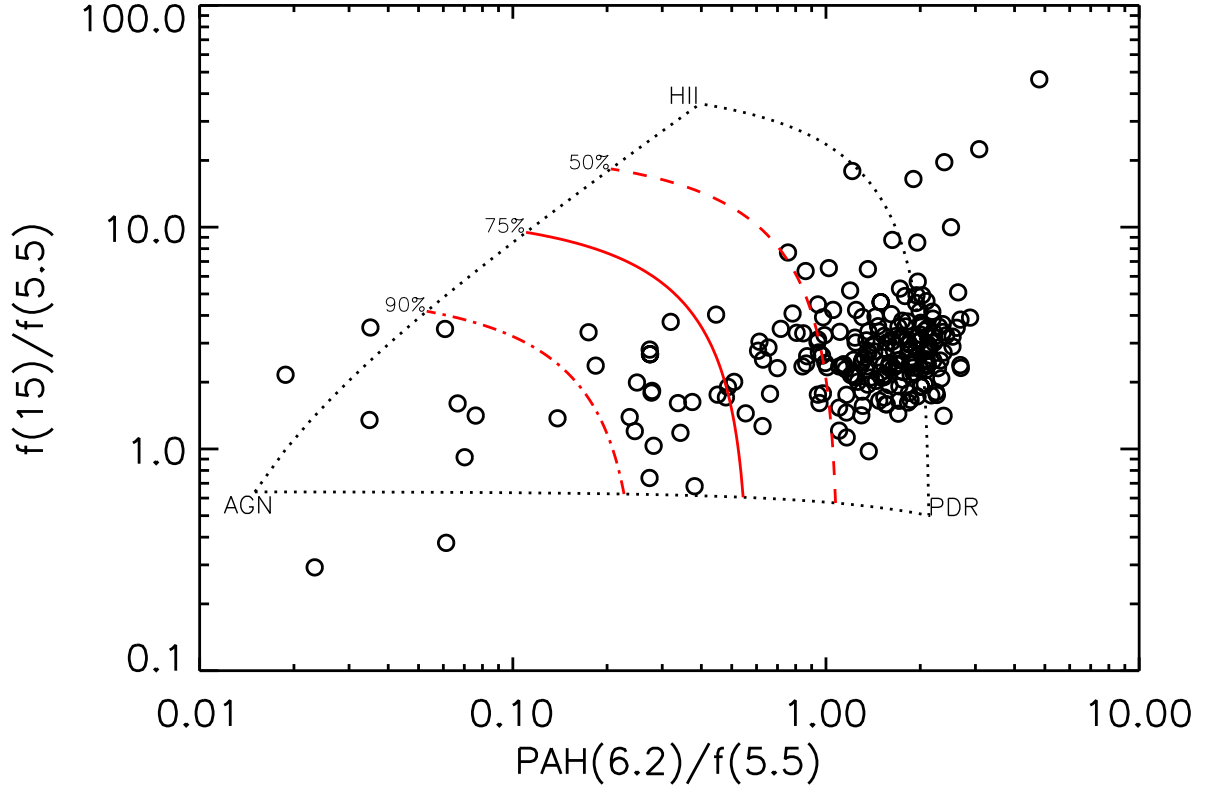}
   \caption{Mid-infrared diagnostic diagram, first used by Laurent et al. (2000) and later modified for Spitzer IRS by Armus et al. (2007), comparing the integrated continuum flux from 14-15 $\mu$m, the integrated continuum flux from 5.3-5.5 and the 6.2 $\mu$m PAH flux. The three vertices, labeled as AGN, HII and PDR, represent the positions of 3C273 from \citet{weedman05}, M17 and NGC 7023 from \citet{peeters04}. These vertices were chosen to facilitate comparison with ULIRGs as presented in \citet{armus07} and \citet{brandl06}.  The red lines from left to right indicate a 90$\%, ~ 75\%$ and 50$\%$ fractional AGN contribution to the nuclear MIR luminosity respectively. GOALs sources are shown as open circles.}
   \label{Laurent-plot}
   \end{center}
   \end{figure}

Figure \ref{Laurent-plot} adopted from Laurent et al. (2000) and Armus et al. (2007) shows the 15 to 5.5 $\mu$m continuum flux ratios versus the 6.2 to 5.5 $\mu$m continuum flux ratios for the LIRGs. The continuum ratios measured in 3C273 \citep{weedman05} were used to represent the expected values for a source in which the MIR luminosity comes only from an AGN, while those of M17 and NGC 7023 \citep{peeters04} typify continuum ratios measured in pure HII and PDR regions. The majority of the LIRGs nuclei fall in the PDR/HII region of the graph, 19\% of objects have low $f_{15 \mu m}/f_{5 \mu m}$ and weak PAH emission suggesting an AGN contribution to the nuclear MIR emission greater than 50\%.

\subsection{Average Spectra}
\begin{figure}[htb]
  \begin{center}
  \includegraphics[width=\linewidth]{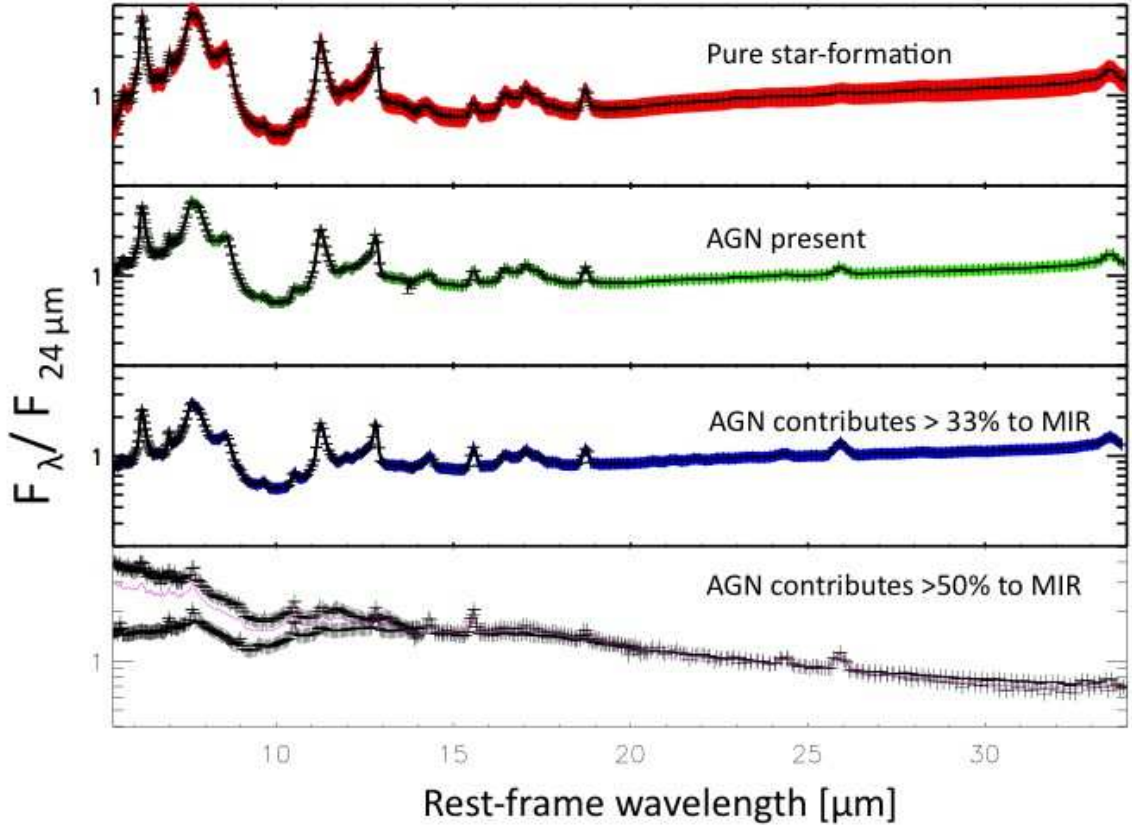}
\caption{Average low-resolution spectra of 4 groups of galaxies. The spectra were normalized by the flux at 24 $\mu$m and were weighted by the signal to noise ratio (SNR) at 24 $\mu$m.  The error-bars give the 1$\sigma$ error on the average while the shaded region shows the intrinsic weighted dispersion in the spectra that were combined to determine the average. From top to bottom we show averages of the following groups of objects: Group (1) contains sources without detectable [NeV] and with [OIV]/[NeII] flux ratios $\leq 0.35$, that is sources whose MIR luminosity is dominated by star-formation. Group (2) contains sources with [NeV] detections. Group (3) contains sources  with [NeV]/[NeII] $\geq$ 0.14 or [OIV]/[NeII]  flux ratios $\geq$ 0.5 suggesting an AGN contribution to the MIR greater than $\sim$ 33\%. Group (4) contains galaxies with [NeV]/[NeII] $\geq$ 0.75 indicating an AGN contribution to the MIR greater than 50\%. Because we have usable SL spectra for only two out of the three galaxies in group (4), instead of showing the intrinsic weighted dispersion on the derived average spectra, we show the actual spectra of the two sources, normalized by the flux at 24 $\mu$m. }

  \label{Avg_spectra}
  \end{center}
  \end{figure}

In order to look for variations in the average spectra of LIRGs as a function of the relative importance of an AGN to the IR emission, we have combined low-resolution spectra, that did not have a large jump between SL and LL, spectra into groups based on detection of the [NeV] 14.3 $\mu$m emission line and the strength of the [NeV]/[NeII] line flux ratio.
Before they were combined, the spectra were normalized to the flux at 24 $\mu$m and were weighted by the signal to noise ratio at 24 $\mu$m. The low resolution spectra were divided into the following 4 groups. The first group contained sources without detectable [NeV] and with [OIV]/[NeII] flux ratios $\leq 0.35$, that is sources whose nuclear MIR luminosity is dominated by star-formation, there are 121 objects in this group. The second group contains sources with [NeV] detections, there are 33 sources in this group. The third group contains all the sources with [NeV]/[NeII] $\geq$ 0.14 or [OIV]/[NeII]  flux ratios $\geq$ 0.5 suggesting an AGN contribution to the MIR greater than 33\% (5 objects). The fourth group contains three sources with [NeV]/[NeII] $\geq$ 0.75 indicating an AGN contribution to the nuclear MIR emission greater than 50\%. For one of the three sources in this group NGC 1068, we were not able to extract usable SL spectra because in the SL map of this source the regions closest to the nucleus suffer from saturation (Howell et al. . The spectra from each group were averaged and are shown in Figure \ref{Avg_spectra}. 

\begin{figure}[htb]
   \begin{center}
   \includegraphics[width=\linewidth]{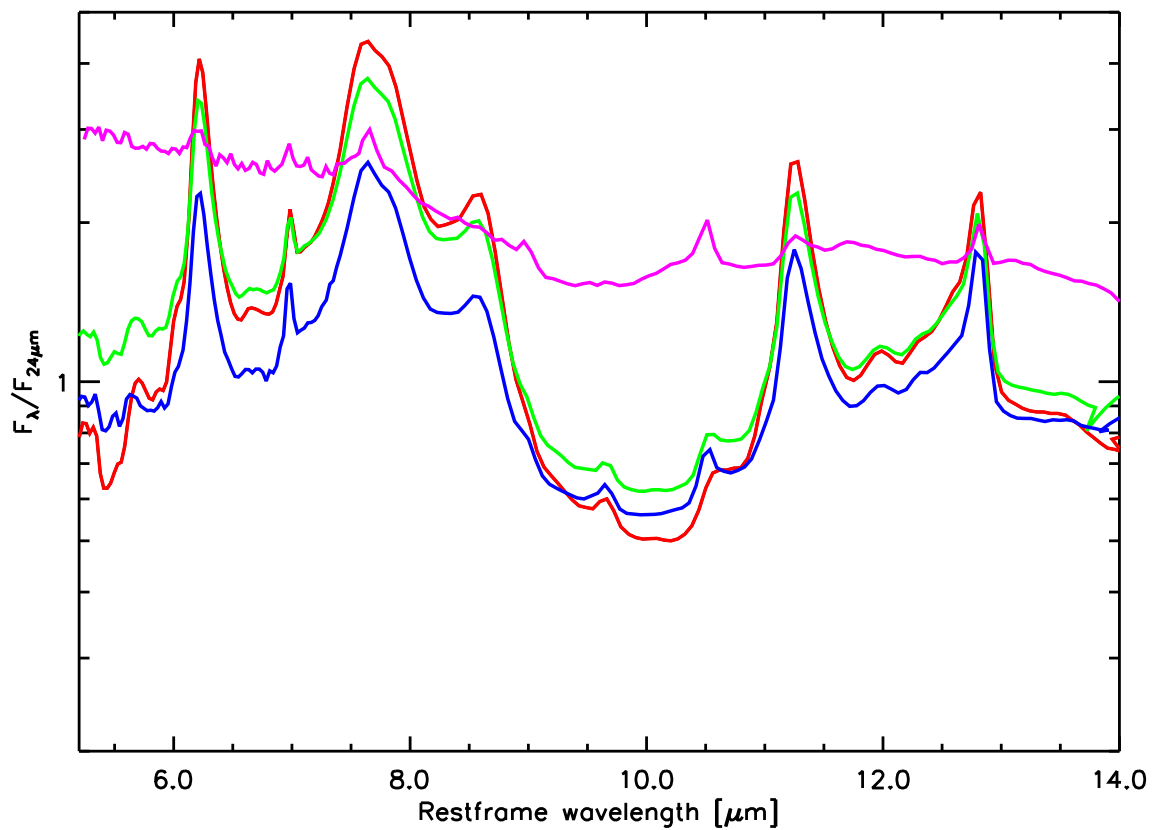}
   \caption{Average spectra of galaxies without detectable [NeV] and with [OIV]/[NeII] flux ratios $\leq$ 0.35 (red), with detectable [NeV]  emission (green), with [NeV]/[NeII] flux ratios $\geq$ 0.14 or [OIV]/[NeII]  flux ratios $\geq$ 0.5 suggesting an AGN contribution to the MIR greater than $\sim$ 33\% (blue), and the average spectra of  two sources in which the AGN dominates the MIR emission (magenta). }
   \label{Zoom-ave}
   \end{center}
   \end{figure}

Qualitatively the average spectra look similar, the broad spectral shapes of both are dominated by PAH emission and silicate absorption. The normalized and scaled comparison of the four average spectra are shown in figure \ref{Zoom-ave}. Given the small number of sources where the [NeV] emission lines indicate that the AGN dominate the MIR emission, we only quantitatively compare groups (1) and (3). The PAH emission, as traced by the 6.2 and 11.3 $\mu$m PAH EQWs is significantly stronger in the group 1 average (EQWs: 0.54 $\mu$m and 0.55 ) than in the average of group 3 objects (EQW: 0.26 and 0.31 $\mu$m). Deeper ice and silicate absorption features are shown in the average spectra of the first group. A statistical analysis of the low resolution spectra of all the GOALS sources will be presented in Stierwalt et al. (2010).    

The correspondence between spectral classification and the continuum slope in the mid-infrared has been discussed by a number of authors \citep[e.g.][] {brandl06, veilleux09a}. The 30 to 15 $\mu$m flux ratio in particular has been used as a powerful diagnostic to separate the AGN from starburst emission in high redshift ULIRGs and QSOs \citep{veilleux09a}. The IR continuum slopes of the average spectra are compared by measuring the 30 $\mu$m to 15 $\mu$m and 30 $\mu$m to 5.5 $\mu$m continuum flux ratios. In the average spectra of group (1) sources, these continuum ratios are 1.6 and 1.8, respectively, while in the average spectra of group (3) sources they both are 1.1, confirming the fact that more hot dust is present in sources with a significant $\geq 33$\% AGN contribution to the MIR. Therefore, the study presented here confirms that, on average, the 30 $\mu$m to 15 $\mu$m and 30 $\mu$m to 5.5 $\mu$m continuum flux ratios can also be used to estimate the relative AGN activity among LIRGs.  

\section{Discussion}
\subsection{AGN contribution to the total IR Luminosity}
In this section we compare the IRS AGN indicators and search for correlations between the AGN contribution to the IR and other properties such as the infrared colors, luminosities and merger stage. 
This paper presents several diagnostics that are effective at isolating the AGN contribution to the MIR emission in LIRG nuclei. The ratios of high ionization [NeV] and [OIV] line fluxes to the low ionization [Ne II] line suggest that only $\sim 1$\% of LIRGs are AGN dominated. The EQWs of the 6.2 $\mu$m PAH feature indicate that 16\% of sources are AGN dominated. Diagnostics based on the shape of the MIR continuum imply that in 19\% of the sources the AGN contributes more than 50\% to the MIR luminosity. 

Several factors may contribute to the apparent discrepancies between the diagnostics presented here. Among these are (1) ambiguities in the definition of the zero point (e.g. the value of the [NeV]/[NeII] and [OIV]/[NeII] ratios for a pure AGN), (2) uncertainties in measuring the continuum under the PAH feature and in measuring the total feature's strength, (3) differences between the projected sizes of the wider SH and LH slits (used to measure the fine structure lines) and the narrower SL slit (used to measure the 6.2 $\mu$m PAH EQW), and (4) differential extinction between [NeV], [OIV] and [NeII] that can affect the measured line ratios. Moreover, inherent in all the MIR diagnostics is the assumption that the bolometric correction is the same for all objects, and that the lines accurately trace either the AGN or the star formation power. A number of authors \citep[e.g.][]{armus07, veilleux09a} have pointed out the failure of these assumptions among ULIRGs and this is expected to be true for LIRGs as well.

While the simple mixing lines of Figures \ref{Nov01_ne5_diag0}, \ref{O4ne2_basic_diag} and \ref{Laurent-plot} give an indication of the AGN contribution to the mid-infrared emission in LIRGs, the real goal is to estimate the fraction of the total bolometric luminosity contributed by an AGN and a SB in each source. To this end, bolometric corrections to the fine-structure lines [NeV], [OIV], [NeII] and hot dust continuum  at $\sim 6 ~\mu$m were applied as in \citet{veilleux09a}. These authors use PG QSO's to estimate the expected  L$_{[\rm{NeV}]}$/L$_{\rm{BOL}}$, L$_{\rm{[OIV]}}$/L$_{\rm{BOL}}$ and L$_{[\rm{NeII}]}$/L$_{\rm{BOL}}$ for sources where the entire IR luminosity comes from an AGN.  We estimate the [NeII] to bolometric (L$_{[\rm{NeII}]}$/L$_{\rm{BOL}}$) ratio for a source with no AGN contribution to the IR emission as the mean of the L$_{[\rm{NeII}]}$/L$_{\rm{BOL}}$ ratio for all sources in this LIRG sample which: (1) have 6.2 $\mu$m PAH EQW $\geq$ 0.54 $\mu$m (2) are not detected in [NeV] nor [OIV] and (3) have [NeV]/[NeII] and [OIV]/[NeII] upper limits of less than 0.1 and 1 respectively. After applying those corrections we estimate the AGN contribution to the nuclear IR luminosity in each source. To determine the AGN contribution to the total IR luminosity we multiply these values by an aperture correction to account for MIR emission outside the slit. It is assumed that the MIR emission outside the slit is produced by young stars and spectral mapping observations of LIRGs confirm this \citep[e.g][]{pere10}. This aperture correction is approximated from the ratio of nuclear to total MIPS 24 $\mu$m flux. It is assumed that this ratio is the same as the ratio of nuclear to total MIR luminosity. At the median distance of the sample (88 Mpc) the width of the SL slit projects to a size of $\sim 1.5$ kpc. The ratio of 24 $\mu$m flux measured with the IRS to the total measured with MIPS has a median of 0.93 and ranges from 0.27 to 1. Thus 
the average aperture corrections to the sample are small. This is especially true for the most luminous galaxies in the sample which are the most distant. 
 After applying these adjustments we find that the whole set of diagnostics converge. We thus estimate that 9-11\% of LIRGs are AGN dominated (i.e. more than 50\% of their IR luminosity is generated by an AGN).

Since large samples of ULIRGs and low luminosity galaxies have been observed with Spitzer, it is interesting to compare the published AGN fractions in those sources to those we have derived for local LIRGs. In ULIRGs 30-40$\%$ of the sources are dominated by AGN dust heating \citep{armus04, armus07, desai07, farrah07, veilleux09a}. In local galaxies with (L$_{\rm{IR}} \leq 10^{11} ~\rm{L}_{\odot}$) this number is closer to 5\% \citet{goulding09}. The data discussed here suggests that LIRGs are powered mostly by star formation and that the AGN contribution to the IR is two times higher in LIRGs than it is in local normal galaxies but three to four times lower than in ULIRGs. The AGN contribution to the total IR luminosity is highest in the most luminous sources (e.g. Armus et al. 2007, Desai et al 2007, Veilleux et al. 2009a, Goulding \& Alexander 2009). 

Figure 6 which shows the IR luminosity versus the 6.2 $\mu$m PAH EQW suggests that within the IR luminosity range of LIRGs there is no evidence of a linear trend between the AGN fraction and the IR luminosity.  Extinction alone cannot explain the different detection statistics in LIRGS and ULIRGs.  While the strength of the silicate absorption cannot be used to determine the differential extinction of [NeV] and [NeII] lines because of the necessary assumptions made about the dust geometry affecting these lines, in most LIRGs the silicate absorption is less deep than in ULIRGs (Pereira-Santanella et al. 2010, Stierwalt et al. in prep). This suggests that ULIRGs may indeed have more extinction in the central few kiloparsecs than LIRGs and the lower AGN fractions among LIRGs cannot be due to extinction alone.

\subsection{Cumulative contribution of AGN to the IR luminosity of the sample}
In the previous section it has been estimated that about 10\% of LIRGs are AGN dominated. 
In this section the total contribution of AGN to the IR luminosity of the entire GOALS sample is computed. 

The cumulative contribution of all AGN to the total bolometric luminosity of the entire sample of local LIRGs was estimated as follows \footnote{As mentioned we refer the values measured within the IRS slit as nuclear values, and total or global values those for the entire galaxy}. (1) The 6.2 $\mu$m PAH EQWs are used to estimate the nuclear AGN contribution to the total IR emission from each galaxy. (2) These values are then multiplied by an aperture correction to account for MIR emission outside the slit as described in the previous section. (3) The AGN IR luminosities for each galaxy are then summed and that sum is divided by the total IR luminosity of the sample. The resulting cumulative contribution of all AGN in the sample to the total IR luminosity of the entire sample of LIRGs is 12\%.


\subsection{AGN fractions as a function of luminosity and color}
\begin{figure}[h] 
$\begin{array}{ccccc}
\includegraphics[width=0.49\linewidth]{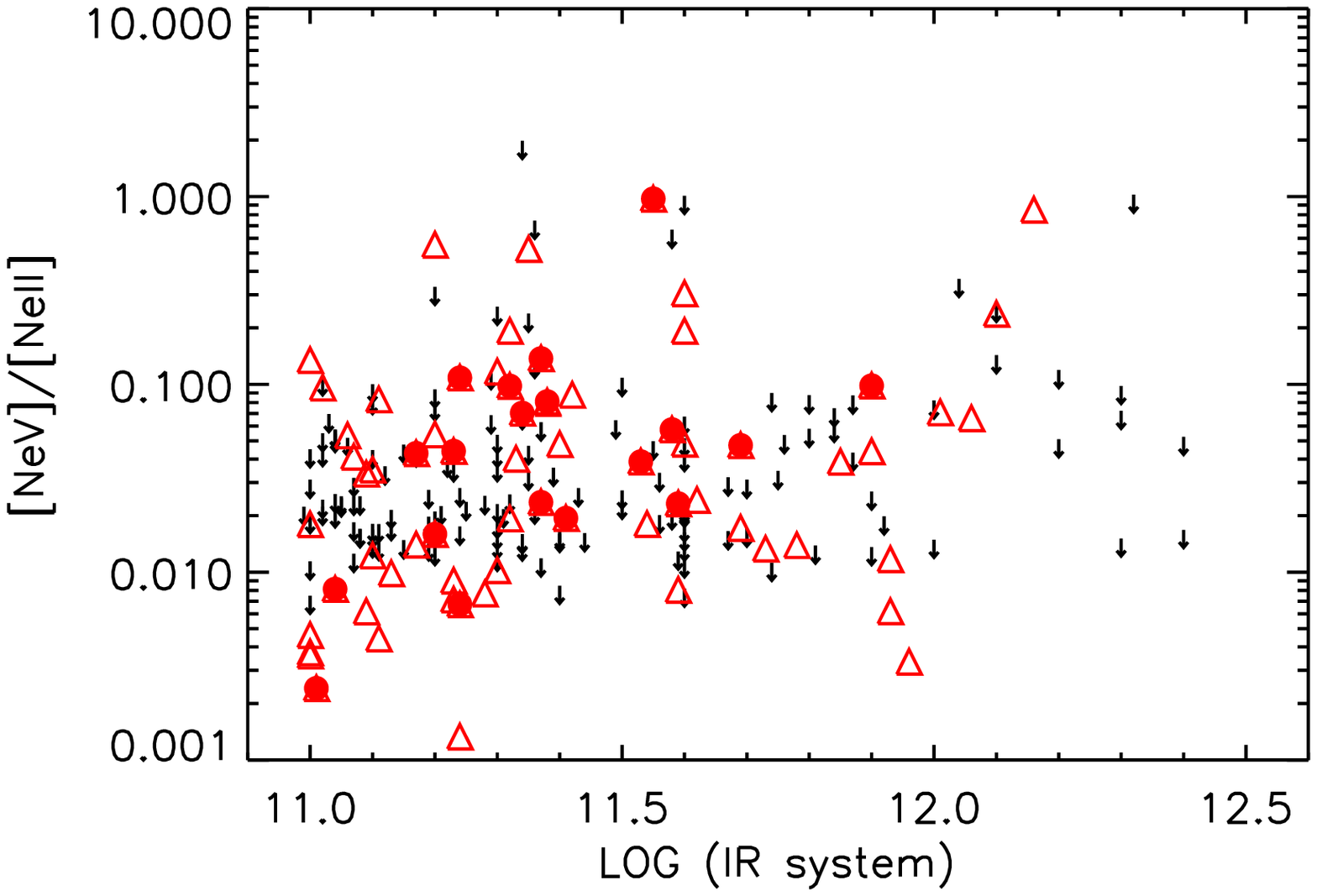}
\includegraphics[width=0.49\linewidth]{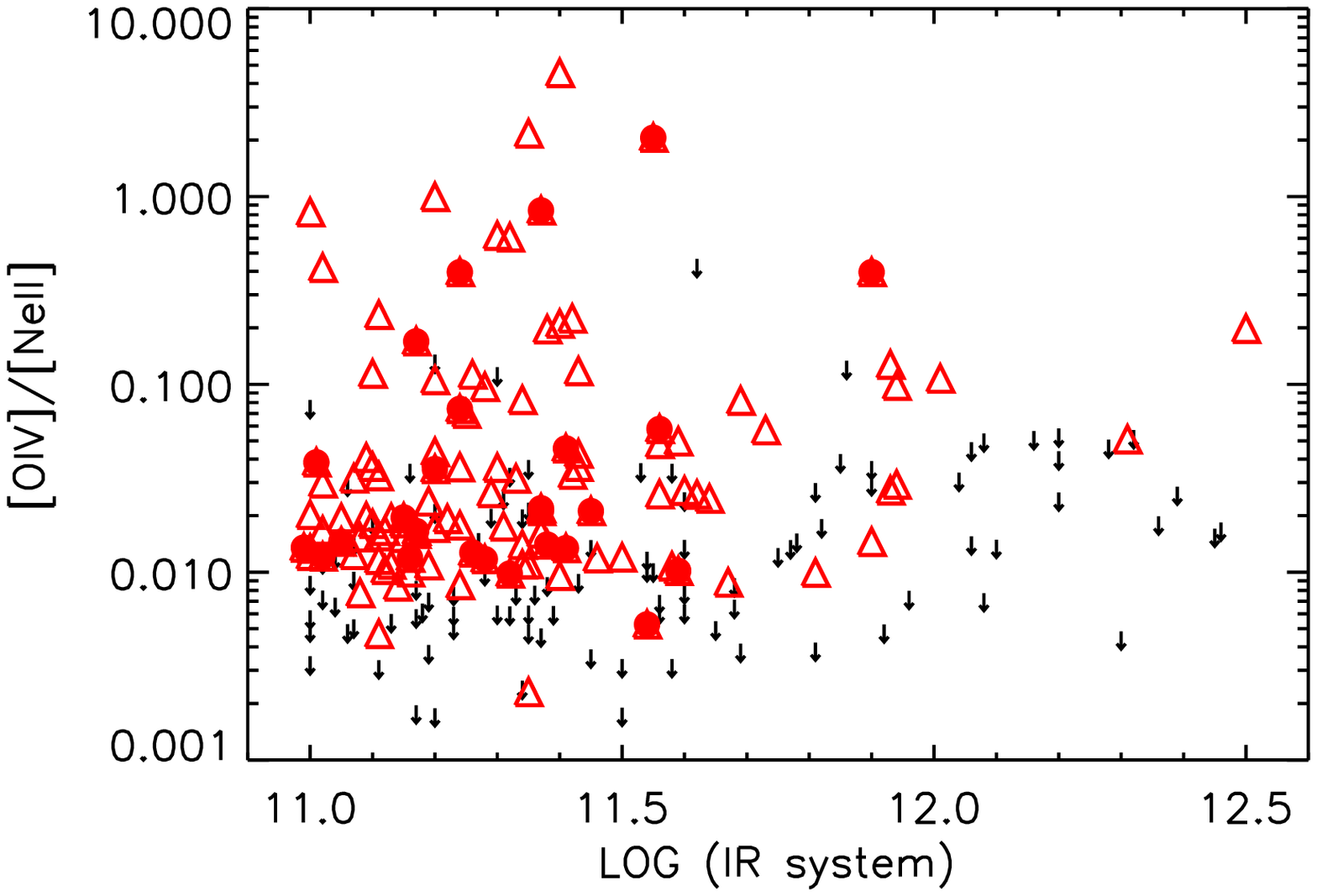}&
\end{array}$
\caption{ Strength of  the [NeV] feature (left) and [OIV] (right)  emission versus the IR Luminosity as estimated from MIPS fluxes (solid symbols) or from IRAS measurements (empty symbols). }
\label{HighResVsIR}
\end{figure}

We find no obvious correlations between the total IR luminosity and the [NeV]/[NeII] and [OIV]/[NeII] emission line ratios (Figure \ref{HighResVsIR}).

\citet{desai07} find that the far-infrared spectral slope is correlated with the AGN contribution in ULIRGs and that despite a large scatter the 6.2 $\mu$m PAH EQW decreases with increasing rest-frame 24 $\mu$m luminosity and with increasing 25 to 60 IRAS flux ratios. Similar relations for LIRGs would permit determining the presence of AGN using only IR photometric studies. 

\begin{figure}[htb]
   $\begin{array}{cc}
   \includegraphics[width=0.49\linewidth]{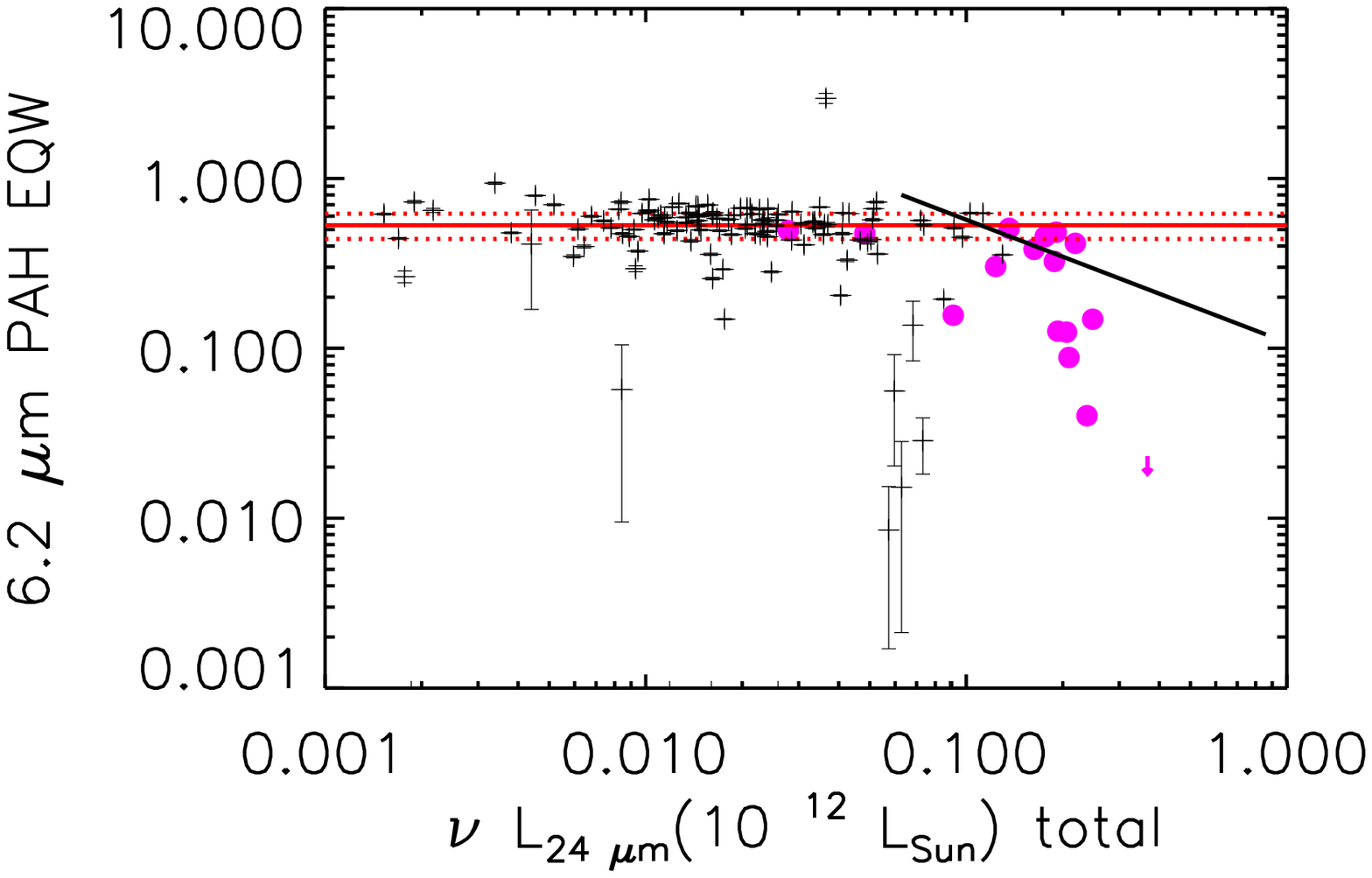}& 
   \includegraphics[width=0.49\linewidth]{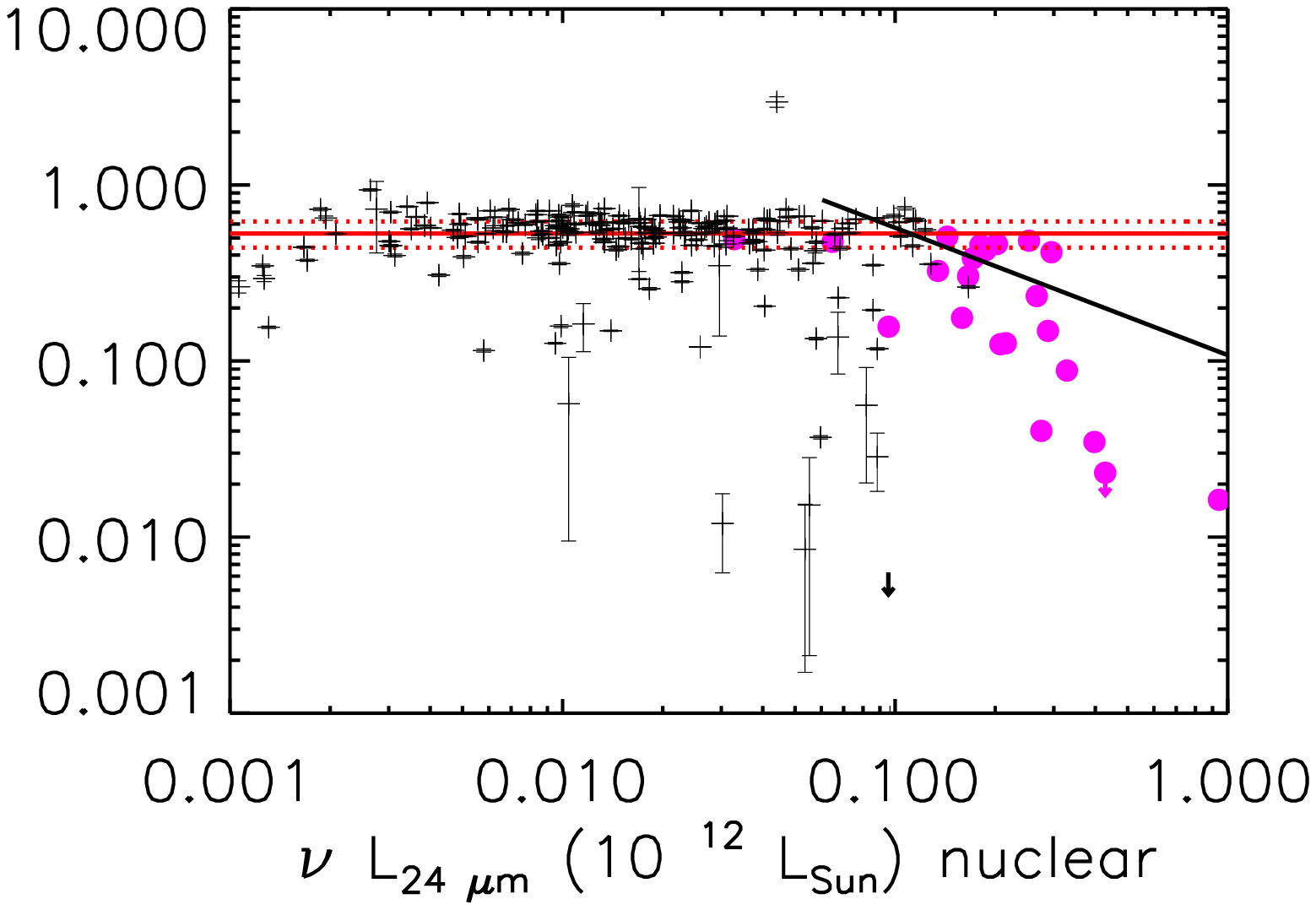}\\
   \end{array}$
   \caption{6.2 $\mu$m PAH EQW versus Log(24 $\mu$m luminosity) from MIPS 24$\mu$m estimates of the luminosity in an aperture of 1\arcmin (left) and from the nuclear LL spectra with a slit width of 10.7\arcsec (right). The 22 ULIRGs from the GOALS sample are shown in magenta filled circles, while the LIRGs are shown as black crosses. The red solid line marks where the 6.2 $\mu$m PAH EQW  equals 0.53 $\mu$m. This is the average EQW for starbursts as determined by \citet{brandl06}. The dotted red lines mark the $1\sigma $ scatter in that value. While no obvious trend with luminosity can be distinguished the median 6.2 $\mu$m PAH EQW  value for LIRGs is higher than that for ULIRGs. The black solid lines marks the relation between the 6.2 $\mu$m PAH EQW versus Log(24 $\mu$m luminosity) found by \citep{desai07}.}
   \label{Lum24-plot}
   \end{figure}

\begin{figure}[htb]
   \begin{center}
   \includegraphics[width=\linewidth]{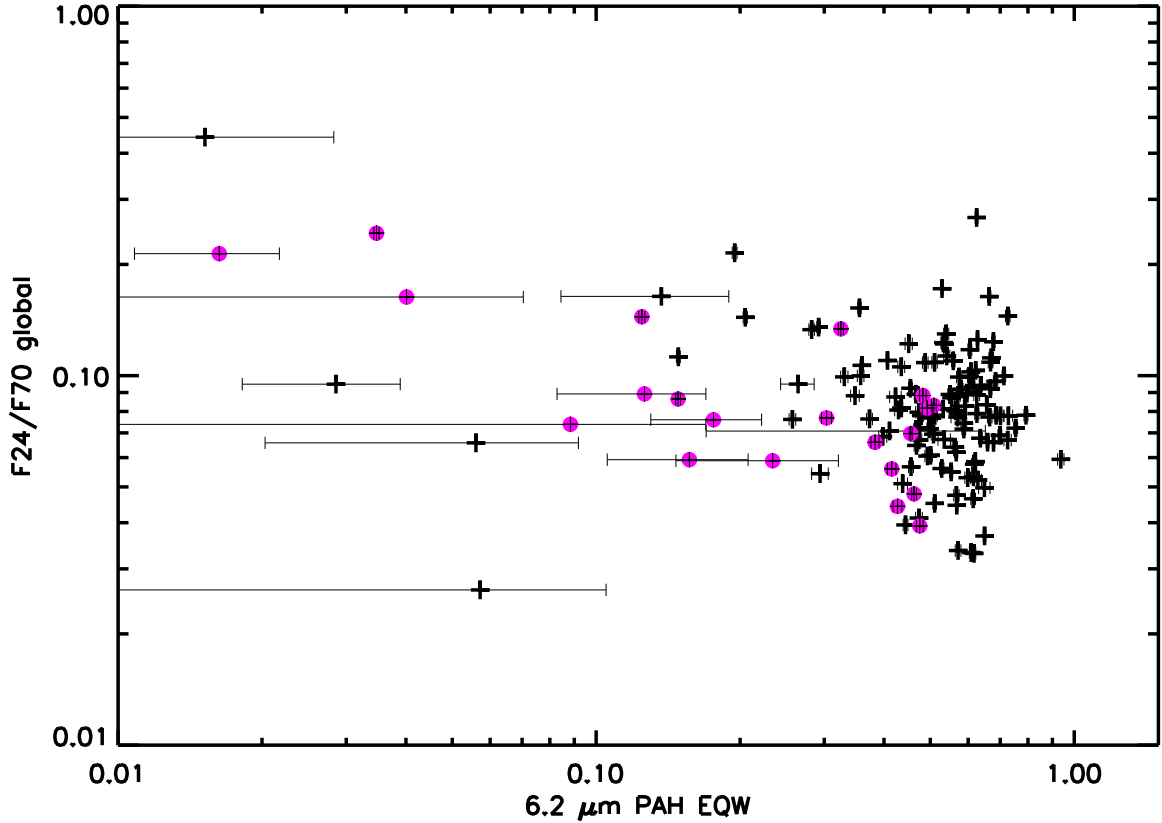}
   \caption{FIR colors versus 6.2 $\mu$m PAH EQW for sources were MIPS 24 $\mu$m and MIPS 70 $\mu$m nuclear fluxes were extracted. The ULIRGs from the GOALS sample are shown in magenta filled circles, while the LIRGs are shown in black crosses. No significant trend with IR colors can be distinguished, however the median MIPS 24 $\mu$m and MIPS 70 $\mu$m flux ratios for AGN dominated sources are higher than those of star-formation dominated sources.} 
   \label{FIRcolor-plot1}
   \end{center}
   \end{figure}

\begin{figure}[htb]
   \begin{center}
   \includegraphics[width=\linewidth]{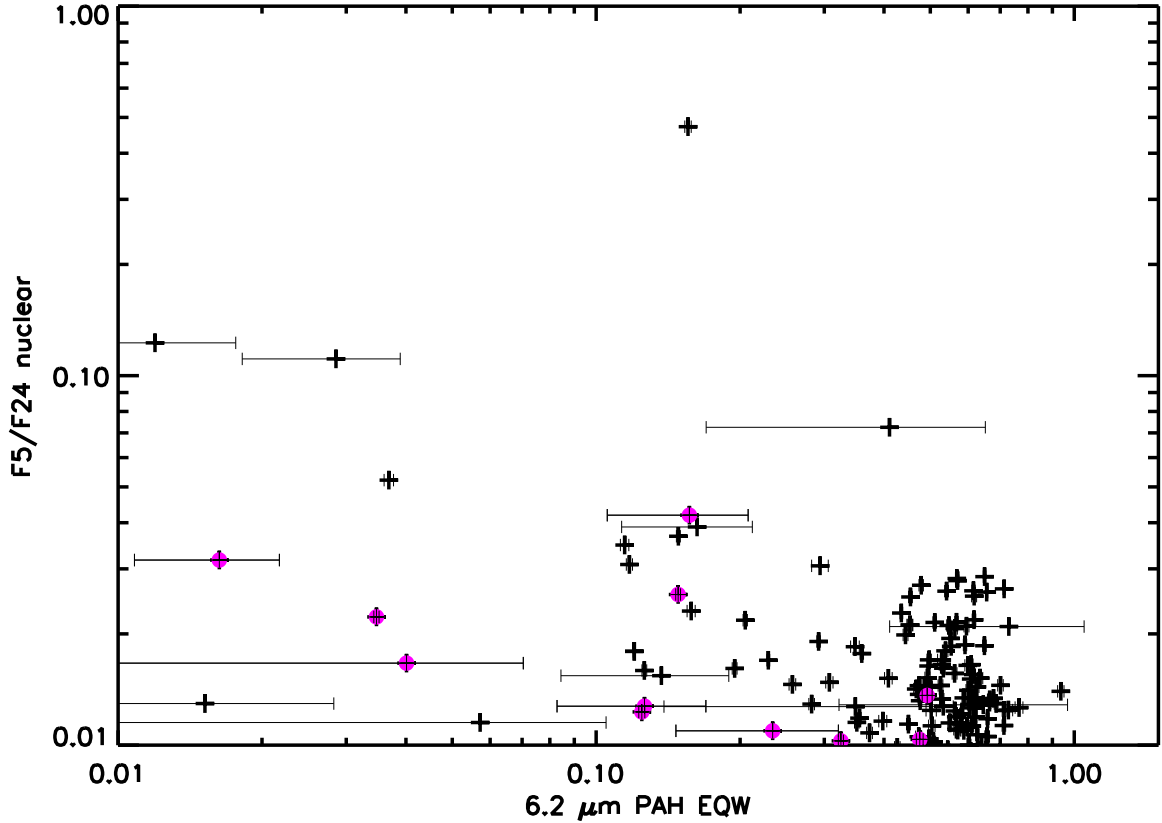}
   \caption{IR colors versus 6.2 $\mu$m PAH EQW. The f5 $\mu$m and f24 $\mu$m nuclear fluxes were extracted from the low resolution spectra. The ULIRGs from the GOALS sample are shown in magenta filled circles, while the LIRGs are shown in black crosses. No significant trend with IR colors can be distinguished, however the median f5 $\mu$m to f24 $\mu$m flux ratios for AGN dominated sources are higher than those of star-formation dominated sources.} 
   \label{FIRcolor-plot2}
   \end{center}
   \end{figure}

Figure \ref{Lum24-plot}, \ref{FIRcolor-plot1}  and \ref{FIRcolor-plot2}  show the distribution of total and nuclear 24 $\mu$m luminosities and $f_{24\mu m}/f_{70 \mu m}$ flux ratios for the GOALS sample versus their 6.2 $\mu$m PAH EQW. Figure \ref{Lum24-plot}, \ref{FIRcolor-plot1} and  \ref{FIRcolor-plot2} show very little scatter in the 6.2 $\mu$m PAH EQW as a function of the total and nuclear 24 $\mu$m luminosity in contrast to the 22 ULIRGs in the sample which show a larger scatter. The median and mean values for $\nu L_{\nu}[10^{12} L_{\odot}]$ for sources with 6.2 PAH EQW ($\leq ~0.27 ~\mu m$) are 0.06 and 0.1 respectively. The same quantities for sources with high 6.2 $\mu$m PAH EQW ($\geq 0.54 ~\mu m$) are 0.02 and 0.02. The range of values for the entire sample is 0.001 to 0.9.

 Figure \ref{Lum24-plot} also shows the distribution of nuclear 24 $\mu$m luminosities obtained from the LL spectra convolved with the response curve of MIPS 24 $\mu$m filter and the results are similar. The mean values of $f_{24\mu m~\rm{MIPS}}/f_{70 \mu m~\rm{MIPS}}$ flux ratios for sources with 6.2 $\mu$m PAH $\leq ~0.27 ~\mu m$ EQW are 0.20 and 0.09 for sources with 6.2 $\mu$m PAH EQW $\geq ~0.54 ~\mu m$. The range of values for the entire sample is 0.02 to 0.88.
 
  A combination of Spearman and Kolmogorov-Smirnov (KS) tests suggests that local LIRGs with significant AGN contribution (6.2 $\mu$m PAH EQW $\leq ~0.27 ~\mu m$)  have higher L$_{24\mu m}$ and are slightly warmer than star-formation dominated sources. The KS test indicates that the flux ratios distributions are different for these two groups of LIRGs with 98\% significance. However, no tight correlation is found between the 6.2 $\mu$m PAH EQW and the nuclear $f_{5 \mu m}/f_{24 \mu m}$ flux ratios (determined from the SL and LL spectra) or the global 24 $\mu$m MIPS / 70 $\mu$m MIPS flux ratios. It should be noted that \citet{desai07} used IRAS colors for the ULIRGs while the analysis presented here is done using MIPS fluxes determined using apertures of 1 arcminute (Mazarella et al. in prep.). However, in local LIRGs IRAS and MIPS fluxes are tightly proportional. The results presented here suggest that IR colors in LIRGs cannot provide an exact estimate of the AGN contribution to the MIR emission in an individual LIRG.  

\subsection{AGN contribution versus stage of interaction}
The connections between galaxy growth through mergers and growth of the central Super Massive Black Hole (SMBH) through accretion have been thoroughly studied theoretically, and are continuously being investigated observationally. Most ULIRGs are advanced stage mergers \citet[e.g][]{murphy00phd, surace98phd}, while local LIRGs span a much wider range of interaction stages. For example many of the lowest luminosity sources ($L_{IR} \sim 10^{11} L_{\odot}$) are single galaxies with only minor companions. Several studies of PG QSOs \citep[e.g.][]{dasyra06, veilleux09b} and warm LIRGs \citet{yuan10} have shown that AGN activity becomes increasingly dominant during the final merger stages. Here we investigate whether this trend is evident among the large sample of local LIRGs in GOALS using MIR spectroscopic diagnostics to determine the presence of an AGN.

\begin{figure}[h] 
$\begin{array}{ccccc}
\includegraphics[width=0.20\linewidth]{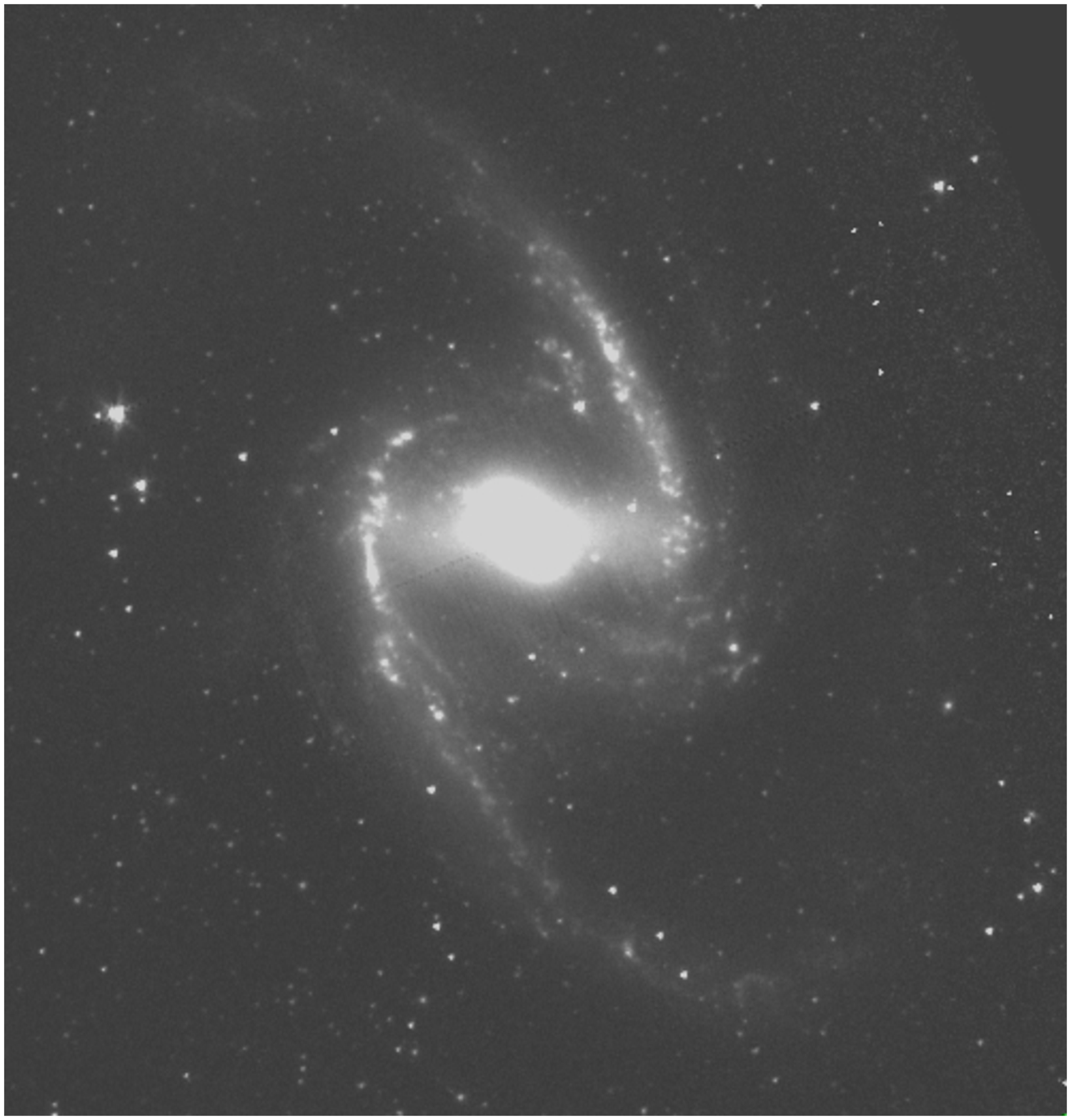}&
\includegraphics[width=0.21\linewidth]{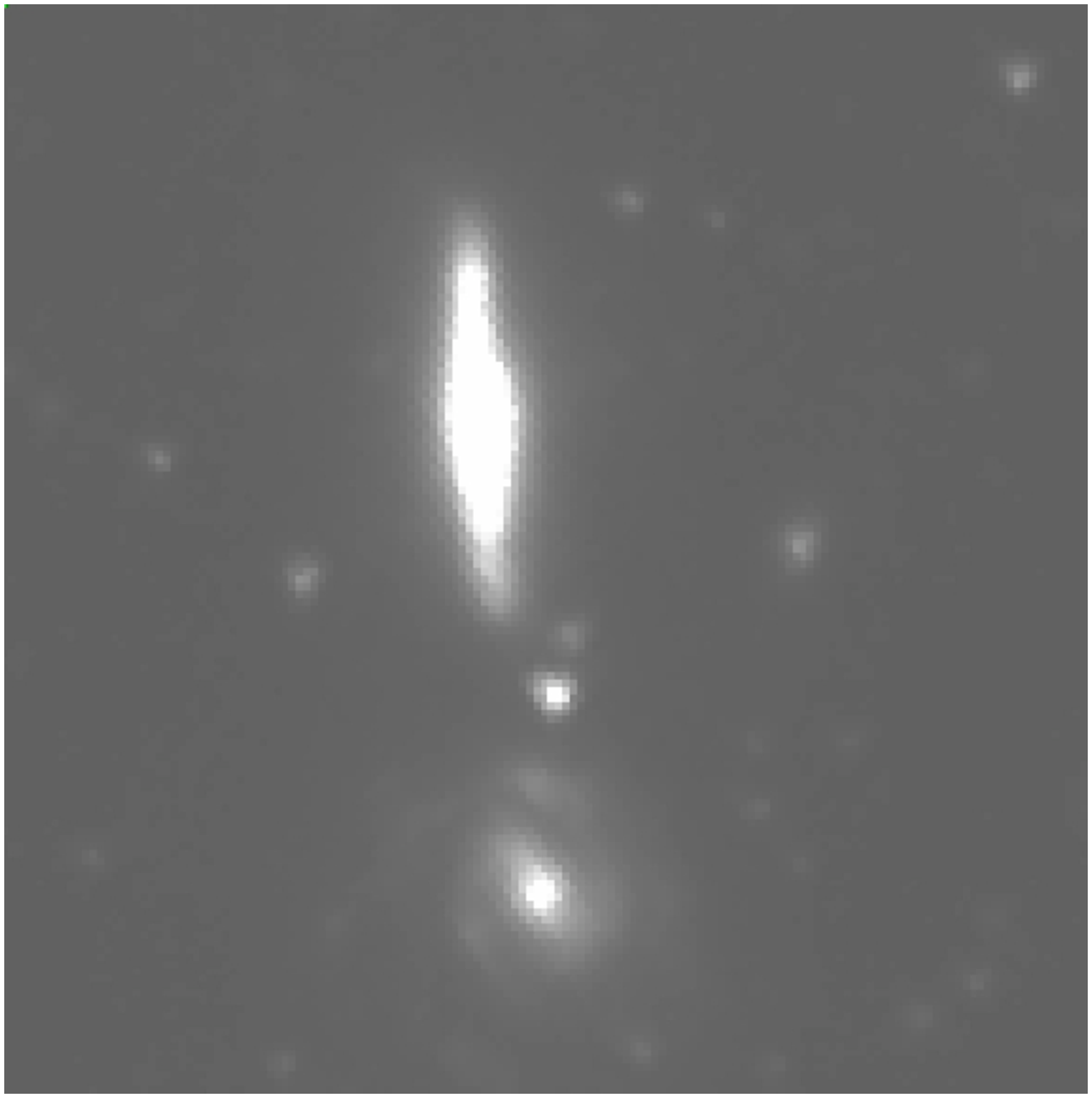}&
\includegraphics[width=0.21\linewidth]{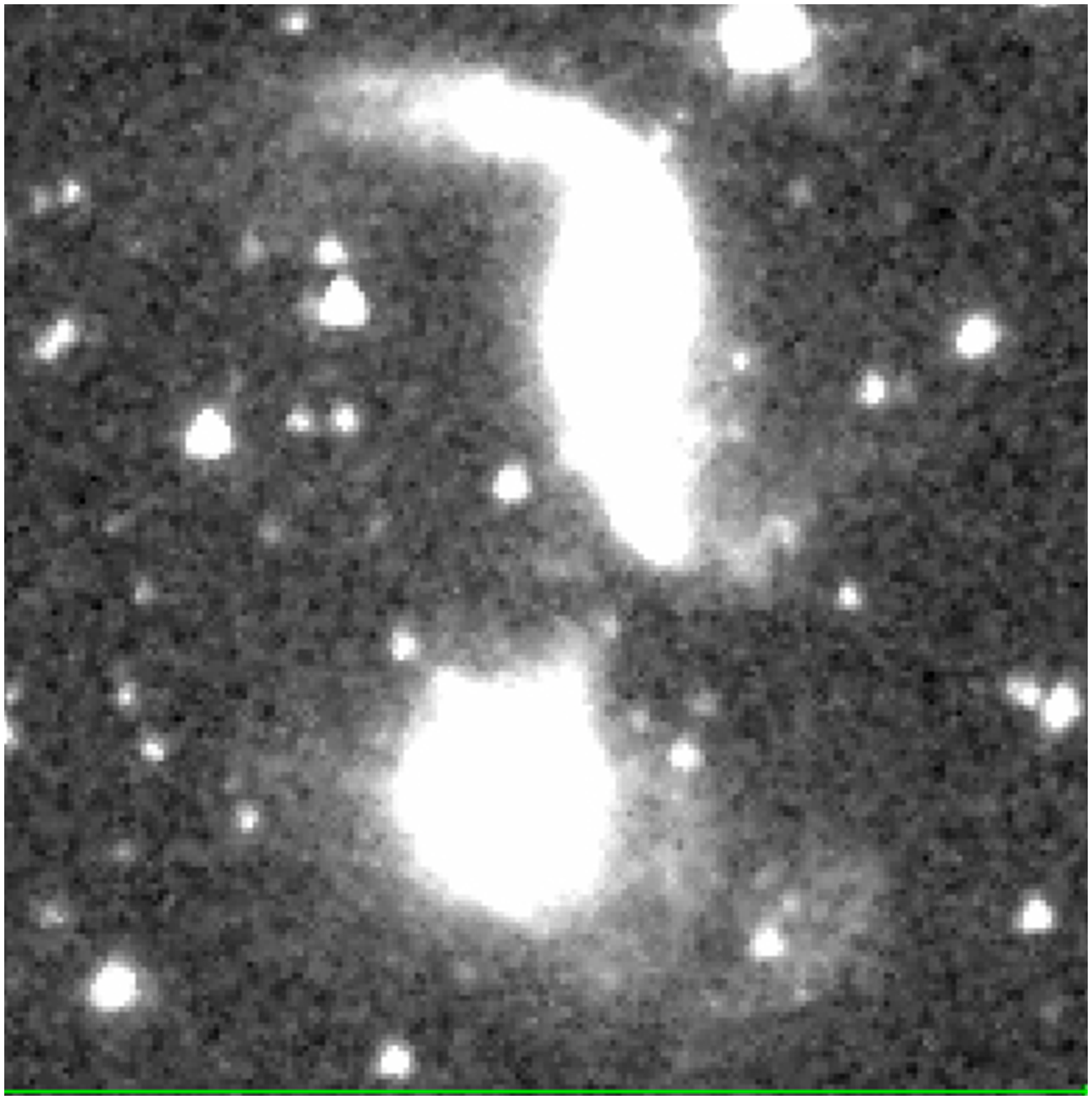}&
\includegraphics[width=0.21\linewidth]{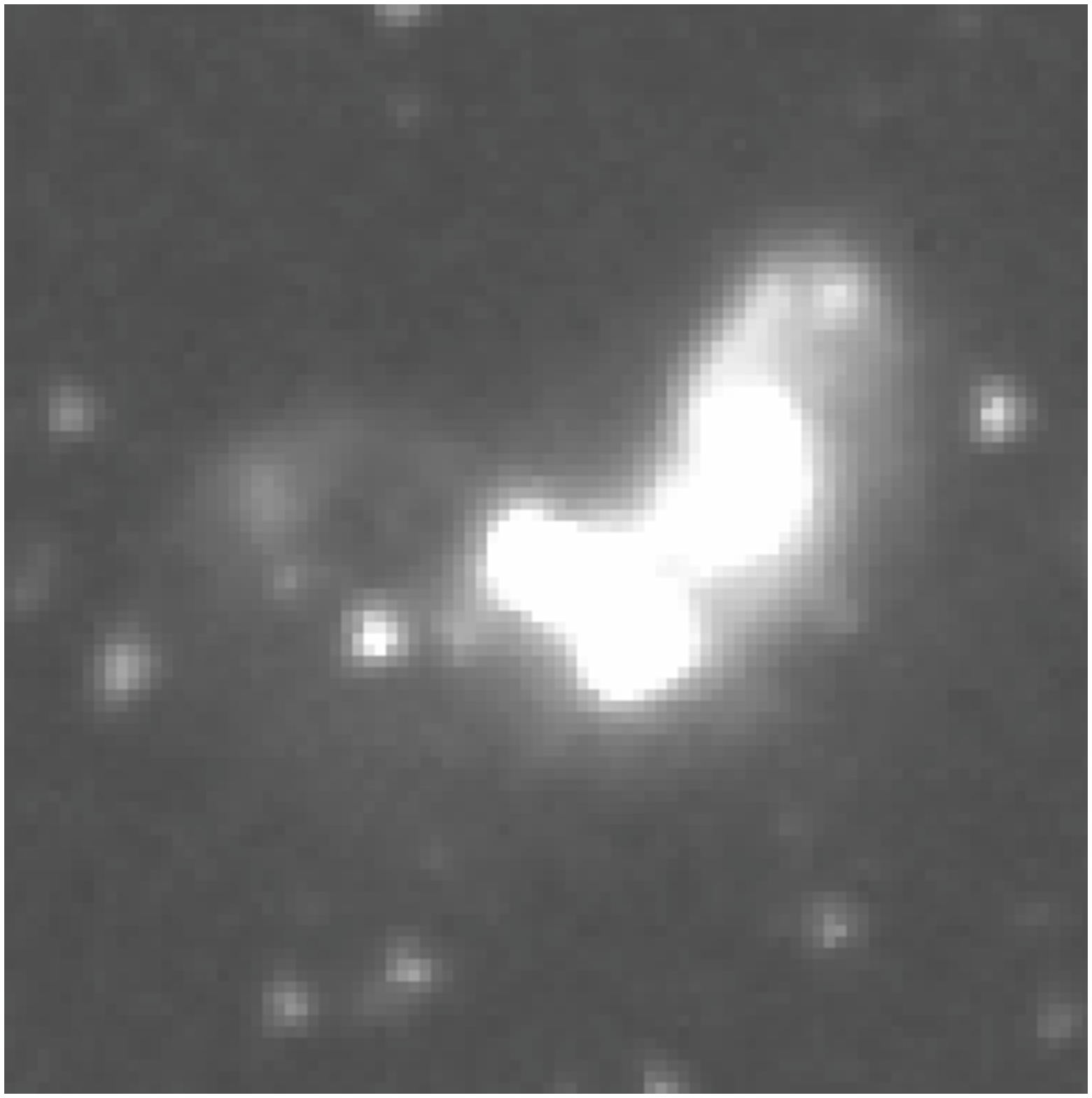}&
\includegraphics[width=0.21\linewidth]{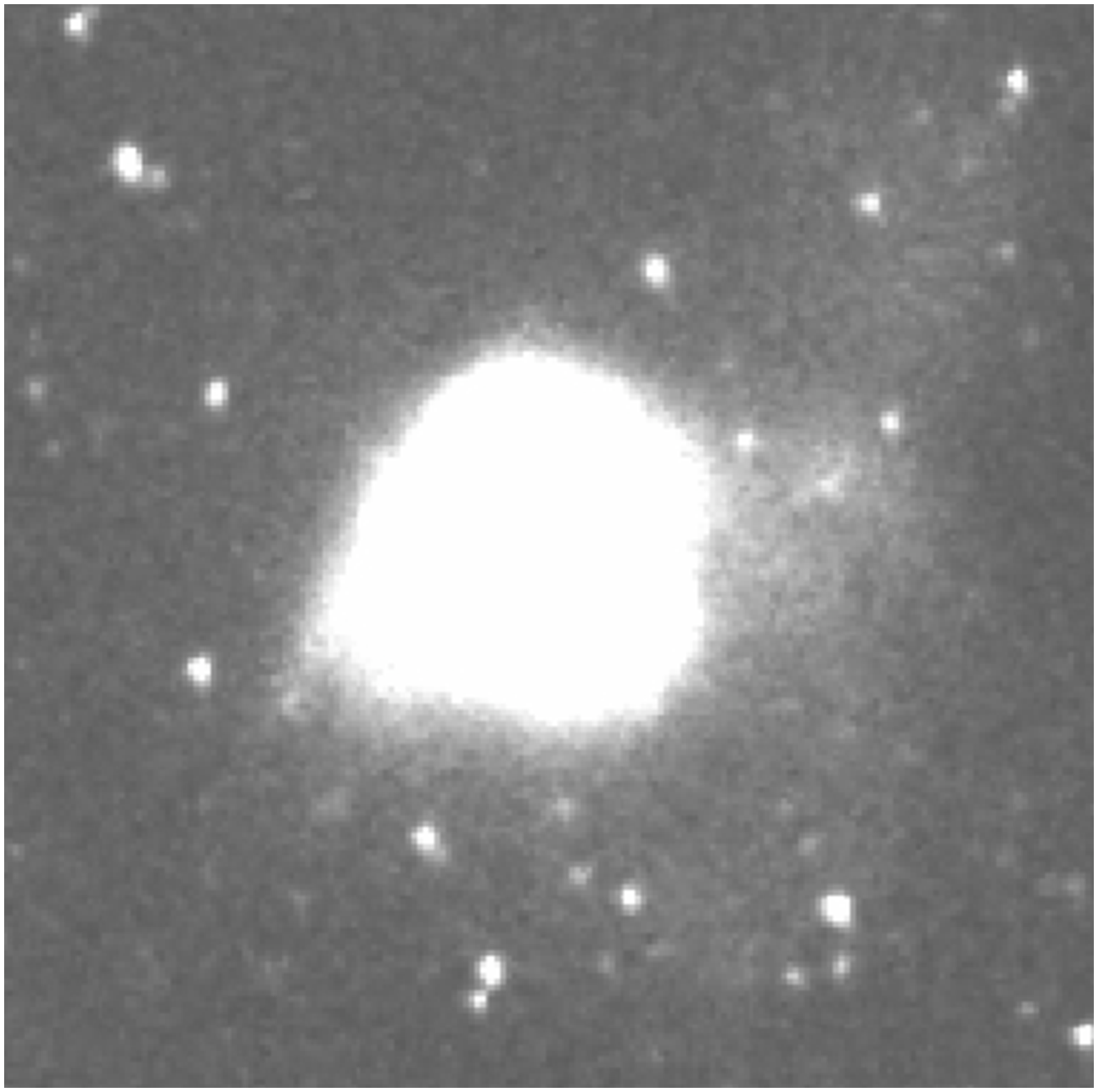}\\
\end{array}$
\caption{IRAC 3.6 $\mu$m images of NGC1365, VV340 (Armus et al. 2009), MCG-02-01-051, IIZw96 (Inami et al. 2010) and Arp 220 illustrating the merger classification stages 0 through 4 used in this paper. Stages are determined as follows(0) no obvious sign of a disturbance either in the IRAC or HST morphologies, or published evidence that the gas is not in dynamical equilibrium (i.e. undisturbed circular orbits), (1) early stage, where the galaxies are within one arc minute of each other, but little or no morphological disturbance can be observed; (2) the galaxies exhibit bridges and tidal tails but they do not have a common envelope and each optical disk is relatively intact; (3) the optical disks are completely destroyed but 2 nuclei can be distinguished; (4) the two interacting nuclei are merged but structure in the disk indicates the source has gone through a merger.  }
\label{mergSeq}
\end{figure}

The GOALS sources were classified in 5 stages (see Figure \ref{mergSeq} for examples): (0) no obvious sign of a disturbance either in the IRAC or HST morphologies, or published evidence that the gas is not in dynamical equilibrium (i.e. undisturbed circular orbits), (1) early stage, where the galaxies are within one arc minute of each other, but little or no morphological disturbance can be observed; (2) the galaxies exhibit bridges, and tidal tails, but they do not have a common envelope and each optical disk is relatively intact; (3) the optical disks are completely destroyed but two nuclei can be distinguished; (4) the two interacting nuclei are merged. The classification scheme was based on a combination of HST, IRAC 3.6, and DSS images. A search in the literature for detailed dynamical analysis based on HI data such as those in Yun et al. (2004) was also done. For the few sources with published HI interferometric data, the merger class was determined on the basis of that data \footnote{For example a source with an HI bridge was classified as stage 2. HI observations are an established method of determining the interaction properties of galaxies and are thoroughly discussed in \citep{hibbard96}.} Note that this paper is not meant to be a comprehensive study of the merger properties of LIRGs. A more detailed study of the multi-wavelength morphologies of LIRGs as well as the effect of extinction on the merger classification will be presented elsewhere. 

\begin{figure}[htb]
   \begin{center}
   \includegraphics[width=\linewidth]{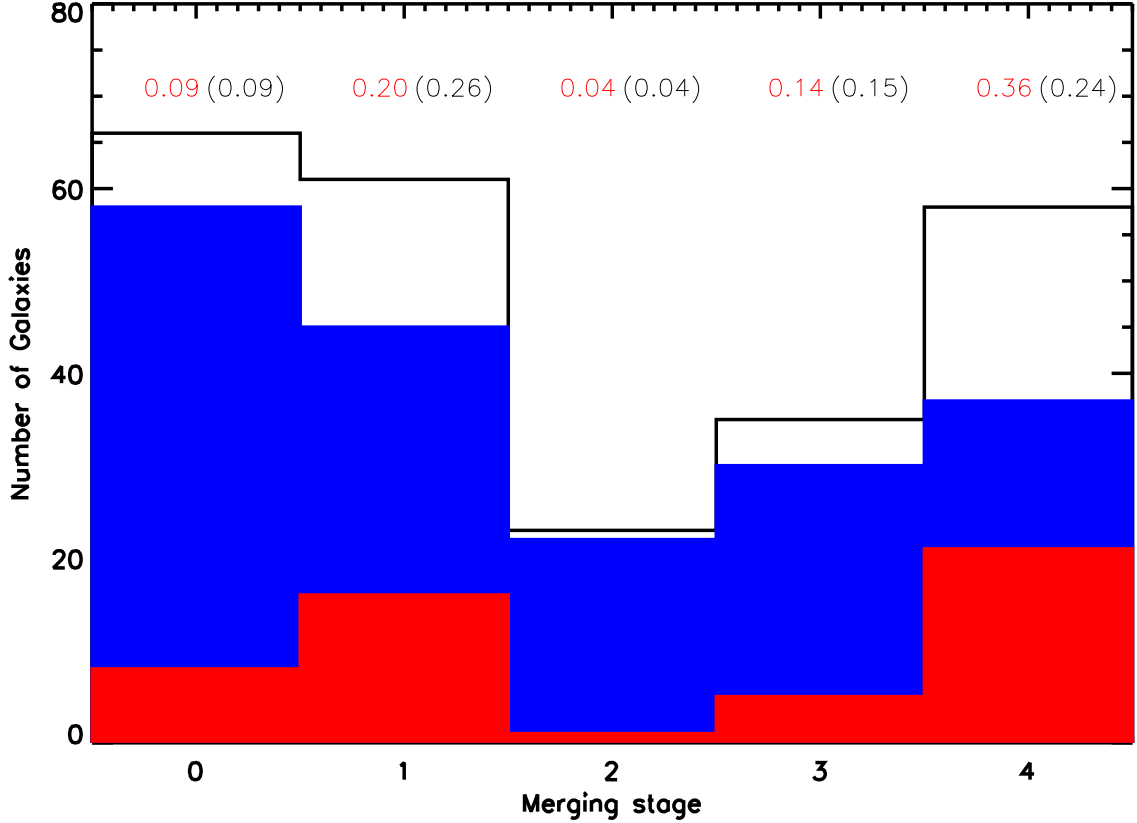}
   \caption{Distributions of AGN dominated sources (i.e. 6.2 $\mu$m PAH EQW $\leq 0.27 \mu$m) in red and starburst objects (i.e. 6.2 $\mu$m PAH EQW $\geq 0.53 ~\mu$m and without detectable [NeV] emission) in blue as a function of merger stage (see section 4.4).  Atop each bin the fraction of AGN dominated sources is written in red together with the same fraction obtained by excluding all ULIRGs (in black in parenthesis). The data is consistent with no trend between the number of AGN dominated sources and the merger stage. However the fraction of AGN dominated sources is significantly higher (40\%) for sources in the last stage of merging. If ULIRGs are excluded from the analysis, the largest fractions (26\% and 24\%) of AGN dominated LIRGs are found in the first and last merger stage respectively.}
   \label{merger-plot}
   \end{center}
   \end{figure}

Figure \ref{merger-plot} shows the histograms of merger stages for AGN dominated galaxies (6.2 $\mu$m EQW $\le 0.27~ \mu $m) and non-AGN sources (sources without detectable [NeV] emission and with 6.2 $\mu$m EQW $\ge$ 0.53 $\mu$m) as well as the fraction of AGN dominated sources as a function of merger stage. We find no strong trends with merger stage. However the fraction of AGN dominated sources is highest for galaxies in the latest merger stages, and the median 6.2 $\mu$m EQW for non-merging galaxies is higher than that for merging systems. A KS test indicates the distribution of 6.2 $\mu$m PAH EQWs of merging GOALS galaxies is different than that of non-merging sources at the 80\% significance level. The last bin in particular is different at the 99\% significance level when compared to the non-merger sources. This appears to be driven both by a decrease in the number of starbursts and by an increase in the number of AGN dominated sources in the final stages of interaction. Sources with double nuclei in stages 1, 2 and 3 are not significantly more AGN dominated than galaxies without obvious evidence of interaction. These results are consistent with models predicting that mergers of gas-rich spirals fuel both star-formation and accretion onto a super-massive black hole, and that the AGN become dominant in producing IR emission in the final mergers stages \citep[e.g][]{sanders88, veilleux09b}. It is interesting to note that this increase in the AGN fraction in the late stage is driven by the ULIRG population. When the 22 GOALS ULIRGs are excluded from this analysis, the highest fractions of AGN dominated sources are found in both early  stage (1) and late stage (4) of interaction. The statistical significance of the difference between mergers and non-mergers in the LIRG only sample decreases to 67\% and the difference between the non-mergers and the last stage sources to 89\%. These results are consistent with the findings of \citep{yuan10} who studied a sample of optically classified LIRGs using optical spectroscopic classifications.

\subsection{IRAC colors}
IR photometry in conjunction with other wavelengths has been used to significantly improve the methodology of finding obscured AGNs \citep[e.g.][]{bark01, glikman04, lacy04, stern05}. AGN exhibit a pseudo power-law continuum in the MIR from dust grains heated to a wide range of temperatures above 100K \citep[e.g.][]{marshall07}. The SEDs arising from each of three distinct sources (AGN, HII region, PDR) which heat the dust are sufficiently different to allow for an AGN selection based on their colors in the range of 4 to 24 $\mu$m \citep[e.g.][]{sajina05, laurent00, lacy07b, glikman04, richards06, stern05, hatzimin05}. 

\begin{figure}[htb]
 \begin{center}
 \includegraphics[width=\linewidth]{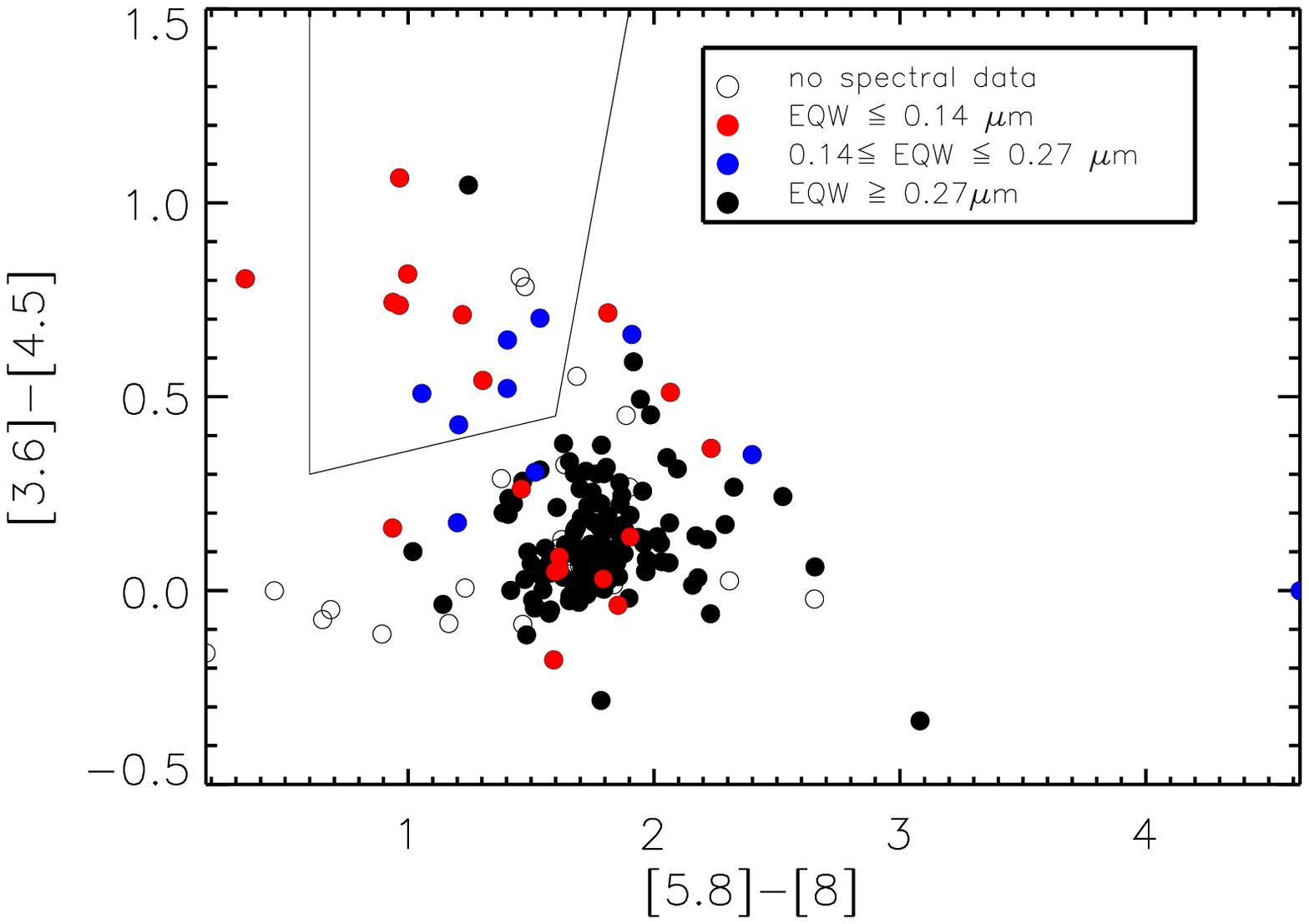}
  \caption{IRAC color ([5.8 $\mu$m ] - [8 $\mu$m]) color ([3.6 $\mu$m ] - [4.5 $\mu$m]) plot of 224 LIRG nuclei. The IRAC total fluxes shown here are from apertures matched to the MIPS 24 $\mu$m apertures 
  of typically 1', but  range between 0.5' to 1.5', (Mazzarella et al. in prep.). The solid lines show the color cuts which Stern et al. (2005) use to separate active galaxies from Galactic stars and normal galaxies. The solid black, red and blue dots represent points with 6.2 $\mu$m PAH EQW greater than  0.27 $\mu$m, between 0.14 and 0.27 $\mu$m and smaller than 0.14 $\mu$m respectively. The empty circles represent nuclei without MIR spectra.}
 \label{irac_plot}
 \end{center} 
  \end{figure}

Figure \ref{irac_plot} shows the IRAC colors for the GOALS sample of LIRGs. All LIRGs chosen to have AGN-like IRAC colors \citep{stern05} also have 6.2 $\mu$m PAH EQW$\leq 0.27 \mu$m indicating a contribution upwards of $\sim$ 50 \% to the IR luminosity. However about half of all sources with low 6.2 $\mu$m PAH EQW fall outside the AGN selection wedge. 
The sources with low 6.2 $\mu$m PAH EQW outside the edge tend to be weaker AGNs as evidenced by (1) only one of those has a weak [NeV] line and (2) their [OIV]/[Ne II] ratios are below 0.2. This suggests that the IRAC color-color method is effective at selecting LIRGs with powerful AGN, but may fail toward the fainter end for LIRGs with an MIR luminosity arising mostly from star-formation. The \citet{lacy04} IRAC color-color method selects 64\% of the AGN dominated LIRGs. However, SB galaxies are also selected as AGN, representing 43\% of all the sources selected with this method. Therefore, for our sample of LIRGs this method finds more AGN but also selects objects that have their IR emission produced by dust heated only by star-formation.


\section{Conclusions}
This paper presents a statistical analysis of 248 LIRG spectra in the rest-frame wavelength range between 5 and 38 $\mu$m. Several diagnostics effective at isolating the Active Galactic Nuclei (AGN) contribution to the Mid-infrared (MIR) emission using [NeV], [OIV] and [NeII] gas lines, the 6.2 $\mu$m PAH EQW and the shape of the MIR continuum are compared. 
In summary:
\begin{enumerate}
\item The high ionization emission lines of [NeV] 14.322 $\mu$m and [OIV] 25.890 $\mu$m are detected in 18\% and 53\% of all LIRG nuclei respectively.
 Since the [NeV] line does not arise from gas heated by hot stars, its detections suggest the presence of an AGN in at least 18\% of LIRG nuclei. 
\item Diagnostics using the [NeV]/[NeII], [OIV]/[NeII] line flux ratios, the 6.2 $\mu$m PAH EQW and the MIR continuum shape suggest that in 10\% of local LIRGs the AGN dominates the bolometric luminosity. The vast majority of LIRGs are SB dominated. The fraction of local LIRG IR emission coming from an AGN as estimated in the mid infrared is approximately two times larger than that seen in normal galaxies  ($\sim$5\%) and, about three to four times lower than that seen in ULIRGs alone.
\item Summing the bolometric luminosity contributed by each AGN in the sample and dividing by the total IR luminosity of all the LIRGs suggests that AGN are responsible for $\sim$12\% of the total bolometric luminosity of local LIRGs.
\item In LIRGs there are no strong correlations between the fraction of IR luminosity from an AGN and the total or nuclear 24 $\mu$m luminosity, the 24 to 60 $\mu$m flux ratios or the interaction stage of the system. However AGN dominated LIRGs tend to be more luminous at 24 $\mu$m and to have warmer IR colors than starburst dominated LIRGs. 
\item By separating the GOALS sources according to merger stage it is found that there is a significant increase in the fraction of AGN dominated sources among those galaxies in the latest stages of interaction. This trend is driven by the ULIRGs in the sample, since these objects tend to be late stage mergers and have larger AGN fractions than the LIRGs . This is consistent with findings of previous authors using optical diagnostics for LIRGs, MIR studies of ULIRGs and PG QSOs and with models which predict that mergers of gas-rich spirals fuel both star-formation and accretion onto a super-massive black hole. 
\item An investigation of the IRAC colors (i.e. [3.6 $\mu$m]-[4.5 $\mu$m] versus [5.8$\mu$m]-[8 $\mu$m]), as introduced in \citet{stern05}, of LIRGs indicates that only 50\% of objects with a significant AGN contribution to the MIR emission fall within the range typically associated with AGN. The \citet{lacy04} AGN IRAC color criteria select a slightly higher fraction of the AGN dominated
LIRGs (64\%), at the expense of also including 11 LIRGs that appear SB dominated from their IRS spectra. 
\end{enumerate}

The measurements we present in this paper provide an estimate of the fraction of IR emission in LIRGs coming from AGN.  Our results provide an important local benchmark against which to compare high-redshift samples of LIRGs, especially at epochs where the contribution of LIRGs to the IR background (e.g. at z$\sim$1 - see Magnelli et al. 2009) becomes substantial.

These diagnostics probe the nuclear source of IR emission in those LIRGs. A full understanding of the processes leading to the generation of LIRG activity requires careful analysis of the mass, temperature and kinematics of the gas fueling and being heated by the star-formation and AGN activity in LIRGs. Future papers (Petric et al. in prep) will discuss the observations of warm and cold molecular gas in the GOALS sample and relate these to the energy sources and evolutionary state of the LIRGs. 

\begin{deluxetable}{lcccccc}
\tableheadfrac{0.0}
\tablecolumns{5}  
\tabletypesize{\scriptsize}    
\tablewidth{0pt}
\tablecaption{[NeV] 14.3 $\mu$m detections in the GOALS sample}
\tablehead{
\colhead{Name}
&\colhead{RA J2000}
&\colhead{Dec J2000}
&\colhead{Flux\tablenotemark{*}  [NeV]}
&\colhead{Error }
&\colhead{${[NeV]}\over{[NeII]}$}
&\colhead{${L_{[NeV]}}\over{L_{IR}}$}\\
\colhead{}
&\colhead{[deg]}
&\colhead{[deg]}
&\colhead{$1.0\times 10^{-18}$}
&\colhead{$1.0\times 10^{-18}$ }
&\colhead{$1.0\times 10^{-2}$}
&\colhead{$1.0\times 10^{-4}$}\\
}
 \startdata
NGC 0232 &	 10.7201   & -23.541 &	38.4 & 5.4 & 31.20 & 1.03 \\
Mrk 1034 &	35.8416    & 32.1971 &	22.6 & 3.3 & 6.31 & 0.33 \\
NGC 1068$^{a}$  &	40.6696 &-0.0133 &	9150.0 & 180.0 & 209.00 & 1.80 \\
UGC 02608 &	48.7561 &42.0357 &	318.00 & 97.1 & 52.5 & 3.67 \\
NGC 1365$^{b}$  &	53.4015  &  -36.1404 &	216.0 & 10.5 & 13.40 & 0.30 \\
ESO 420-G013 & 63.4571   & -32.0070 &	113.0 & 7.9 & 9.60 & 0.76 \\
UGC 03094 &	 68.8910   &  19.1717&	15.3 & 3.0 & 4.79 & 0.20 \\
CGCG 468-002NED01 &77.0821    & 17.3633&	21.3 & 1.8 & 28.10 & 0.54 \\
IRAS F05081+7936 &	79.1933 &79.6703 &	22.6 & 3.7 & 4.40 & 0.46 \\
IRAS F05189-2524$^{c}$ &	80.2559 & -25.3626 &	151.0 & 11.8 & 85.00 & 1.06 \\
IRAS 05083+244 &	77.8578 &24.7551 &	11.3 & 3.9 & 1.28 & 0.18\\ 
IRAS F07027-6011 &	105.8506 &-60.2561 &	26.9 & 2.1 & 19.30 & 0.44 \\
IRAS 16164-0746 &	244.7991    &-7.9008&	11.9 & 2.9 & 2.40 &0.18\\  
NGC 2623 &	129.6003    & 25.7547 &	26.2 & 4.0 & 4.82 & 0.13 \\
2MASX J09133644-1019296 &138.4021   & -10.3249 &	12.5 & 1.6 & 9.84 & 0.23 \\
2MASX J09133888-1019196 &138.4120   & -10.3221 &	13.8 & 1.6 & 10.80 & 1.17 \\
UGC 5101$^{d,f}$ &	143.9652   &  61.353 &	 28.6 & 1.9 & 7.08 & 0.22\\
ESO 267-G030 & 183.5534  &  -47.2285 &	19.4 & 3.7 & 4.41 & 0.28 \\
NGC 4922NED02 &	195.3553   &  29.3138 &	30.8 & 6.4 & 8.10 & 0.41 \\
UGC 08387 &	200.1473  &   34.1395  &	14.8 & 4.4 & 1.33 & 0.09 \\
MCG -03-34-064 &	200.6019  &  -16.7284 &	547.0 & 19.7 & 112.00 & 6.55 \\
NGC 5135$^{e}$ &	201.4332 &-29.8333 &	 106.0 & 6.3 & 11.70 & 0.45 \\
Mrk 266$^{b}$  & 	204.5737 & 48.2761 & 	21.5 &  2.9 &  14.05 & 0.28 \\  
NGC 5256 & 	204.5719 & 48.2756 & 	79.6 &  12.8 &  13.70 &  1.42 \\
Mrk 273$^{b}$  & 	206.1755 & 55.8870 & 	101.0 &  3.3 &  23.70 &  0.66 \\
NGC 5734 & 	 221.2959   & -20.9135 & 	9.9 &  3.3 &  2.90 &  0.10 \\
VV340a & 	224.2529   &  24.6183 & 	13.3 &  2.1 &  4.74 &  0.16 \\
NGC 5990 & 	236.5684 & 2.4154 & 	12.0 &  4.1 &  1.94 &  0.08 \\
NGC 6156 & 	248.7190 & -60.6189 & 	11.9 &  4.2 &  4.11 &  0.06 \\
NGC 6240$^{f}$ & 	253.2454 & 2.4009 & 	20.7 &  7.2 &  1.17 &  0.07 \\
CGCG 141-034 & 	269.2360 & 24.0172 & 	20.3 &  1.8 &  5.40 &  0.28 \\
CGCG 142-034B & 	274.1410  &   22.1108 & 	10.1 &  2.7 &  8.30 &  0.15 \\
ESO 339-G011 &  299.4067  &  -37.9357 & 	245.0 &  4.1 &  55.30 &  3.06 \\
MCG +04-48-002 & 	307.1461  &   25.7334 & 	29.0 &  5.6 &  5.34 &  0.28 \\
NGC 6926 & 	308.2755   &  -2.0275 & 	13.4 &  2.4 &  19.30 &  0.13 \\
NGC 7130 & 	327.0813   & -34.9517& 	58.6 &  4.6 &  8.79 &  0.27 \\
NGC 7469 & 	345.8151     & 8.8740 & 	119.0 &  15.4 &  5.75 &  0.38 \\
NGC 7679 & 	352.1943    &  3.5115  & 	32.9 &  4.2 &  5.2 &  0.39 \\ 
CGCG 453-062 & 	 346.2355    & 19.5522 & 	22.7 &  6.9 &  7.95 &  0.29 \\
ESO 148-IG002 & 	348.9459   & -59.0547 & 	20.4 &  3.9 &  6.59 &  0.19 \\
IC 5298 & 	 349.0029    & 25.5567& 	105.0 &  4.9 &  30.50 &  0.79 \\
NGC 7592 & 	349.5946   &  -4.4162 & 	14.9 &  1.5 & 3.60 &  0.21 \\   
NGC 7674 & 	351.9863 & 8.7790 & 	186.0 &  7.8 &  97.30 &  2.16 \\
	                \enddata
               
  \tablenotetext{*}{Notes: All fluxes are given in W m$^{-2}$. The fluxes presented here agree with previously published values within the errors, in all cases with the exception of fluxes for Mrk 266, NGC5135. }
  
  \tablenotetext{a}{Howell et al. 2007}
  \tablenotetext{b}{Dudik et al. 2007}
 \tablenotetext{c}{Armus et al. (2007)}
  \tablenotetext{d}{Farrah et al. (2007)}
  \tablenotetext{e}{Gorjian et al. (2007)}
  \tablenotetext{f}{Armus et al. 2006}
  \end{deluxetable}




%
%
%
%

{\it{Acknowledgments:}} We thank the anonymous referee for his comments which have significantly improved our paper. VC acknowledges partial support from the EU grants ToK 39965 and FP7-REGPOT 206469. AP thanks N. Flagey for multiple readings of the document comments which helped improve and clarity of this text. AP also thanks V. Desai for help with the data analysis and C. Bridge for discussions and help with the merger classification of the LIRGs in this paper. 

This work is based primarily on observations made
with the Spitzer Space Telescope, which is operated
by the Jet Propulsion Laboratory, California Institute
of Technology under NASA contract 1407.

We have made use of the NASA/IPAC Extra-
galactic Database (NED) which is operated by the Jet
Propulsion Laboratory, California Institute of Technol-
ogy, under contract with NASA. Support for this re-
search was provided by NASA through an award issued
by JPL/Caltech. 

\bibliographystyle{aa} 

\end{document}